\documentclass[aps,prb,twocolumn,longbibliography,showpacs,superscriptaddress,amsmath,amssymb]{revtex4-1}

\newcommand{\bra}[1]{\langle #1|}
\newcommand{\ket}[1]{|#1\rangle}
\newcommand{\braket}[2]{\langle #1|#2\rangle}

\usepackage{graphicx}  
\usepackage{bm}
\usepackage{xcolor}
\usepackage{hyperref}

\begin{document}

\title{Excitons in boron nitride single layer}
\author{Thomas Galvani}
\author{Fulvio Paleari}
\author{Henrique P. C. Miranda}
\author{Alejandro Molina-S\'{a}nchez}
\author{Ludger Wirtz}
\affiliation{Physics and Materials Science Research Unit, University of Luxembourg, 162a avenue de la Fa\"iencerie, L-1511 Luxembourg, Luxembourg, EU}
\author{Sylvain Latil}
\affiliation{CEA, IRAMIS, SPEC, GMT, 91191 Gif-sur-Yvette, France}
\author{Hakim Amara}
\affiliation{Laboratoire d'Etude des Microstructures, ONERA-CNRS, BP 72, 92322 Ch\^atillon Cedex, France}
\author{Fran\c cois Ducastelle}
\affiliation{Laboratoire d'Etude des Microstructures, ONERA-CNRS, BP 72, 92322 Ch\^atillon Cedex, France}

\date{\today}

\begin{abstract}
Boron nitride single layer belongs to the family of 2D materials whose optical properties are currently receiving considerable attention. Strong excitonic effects have already been observed in the bulk and still stronger effects are predicted for single layers. We present here a detailed study of these properties by combining \textit{ab initio} calculations and a tight-binding-Wannier analysis in both real and reciprocal space. Due to the simplicity of the band structure with single valence ($\pi$) and conduction ($\pi^*$) bands the tight-binding analysis becomes quasi quantitative with only two adjustable parameters and provides tools for a detailed analysis of the exciton properties. Strong deviations from the usual hydrogenic model are evidenced. The ground state exciton is not a genuine Frenkel exciton, but a very localized ``tightly-bound'' one. The other ones are similar to those found in transition metal dichalcogenides and, although more localized, can be described within a Wannier-Mott scheme.
\end{abstract}


\maketitle

\section{Introduction}
Two-dimensional materials are currently the object of many investigations concerning their electronic and optical properties. Graphene is the most known example\cite{Castro2009} but hexagonal boron nitride,\cite{Watanabe2006,Jaffrennou2007,Watanabe2009,Museur2011,Pierret2014} and now transition metal dichalcogenides (TMD)\cite{Zhao2012,Xu2013,Zhang2015,Molina2015,Wu2015} as well as new materials such as phosphorene,\cite{Ling2015,Favron2015} silicene, germanene, etc.\cite{Vogt2012,Li2014} are receiving considerable attention. Apart from graphene these materials are semiconductors which new optical properties compared to those of the familiar 3D semiconductors. Excitonic effects, in particular, are  more pronounced in 2D than in 3D, with the exciton binding energies being much higher, of the order of 0.1 -- 1 eV or more. The spatial extension of the excitons remains fairly large in general so that they are frequently considered as Wannier-Mott excitons. However, it has been quickly noticed  that the usual hydrogenic model does not apply in 2D because of the  different screening processes involved.\cite{Mak2010,Splendiani2010,Cudazzo2011,Chernikov2014,Huang2013,Molina2013,Qiu2013,He2014,Pulci2015} There is thus a need to understand more precisely these excitonic effects. The case of hexagonal boron nitride (hBN) is more specific still. Even in its bulk hexagonal form, very strong excitonic effects have been reported very early based on ab-initio calculations.\cite{Arnaud2006} The theoretical interpretation of a very strongly bound excitons (0.7 eV for the ground-state exciton in bulk hBN) was confirmed by various experiment\cite{Watanabe2006,Jaffrennou2007,Watanabe2009,Museur2011,Pierret2014} and was refined in theoretical calculations considering symmetry arguments.\cite{Wirtz2006,Wirtz2008,Arnaud2008} The reason for the strong binding energy is the quasi-2D nature of the hBN structure\cite{Arnaud2006} consisting of stackings of hexagonal layers interacting through weak (mostly Van der Waals like) interactions. Furthermore hBN has a very large gap, $>$ 6~eV, leading to a rather weak dielectric screening, so that all ingredients conspire to enhance these excitonic effects. They have been studied recently, but the experiments are difficult because of the necessity to work in the far UV range.

The current interest in 2D materials and the development of techniques to handle few-layer materials suggest of course to study the properties of hBN as a fonction of the number of layers, as has been done in the case of graphene and TMD. In the case of TMD it has been shown that the nature of the gap, indirect or direct, depends on this number of layers. What about hBN ? Preliminary experimental studies are already available,\cite{Schue2015} but much remains to be done, and first, a precise knowledge of the single layer (SL) properties is required.

This is the main purpose of the present article. We present a detailed theoretical study of the first excitonic levels, and characterize their energies and shape by combining \textit{ab initio} calculations and a simple tight-binding model. The \textit{ab initio} approach is the usual one, based on a GW plus Bethe-Salpeter approach. The tight-binding approach is close to the approach put forward by Wannier long ago.\cite{Wannier1937,Knox1963,Toyozawa2003,Bechstedt2015} It turns out here that we have basically to take into account just one $\pi$ valence band and one $\pi^*$ conduction band. Furthermore, close to the gap, the corresponding Bloch states are concentrated on the nitrogen (N) and boron (B) atoms respectively, so that the usual $\pi$ orbitals can be considered as genuine Wannier functions. It is then possible to work out the Wannier equations in real space in a fairly simple but surprisingly accurate way.

The paper is organised as follows: Section \ref{section_estructure}  is devoted to the electronic structure of hBN-SL  which is calculated using standard \textit{ab initio} techniques and fitted to a simple tight-binding model. Section \ref{section_excitons} contains the principal discussion of the various excitons and of their symmetry, using in particular imaging tools in real and reciprocal spaces. Finally Section \ref{optics} is concerned with the calculation of optical matrix elements. This is followed by a discussion (Section \ref{discussion}) and several appendices.

\section {Electronic structure of  \lowercase{h}BN single layer}
\label{section_estructure}
\subsection{Band structure}

\begin{figure}[!ht]
\begin{center}
\includegraphics*[width=7cm]{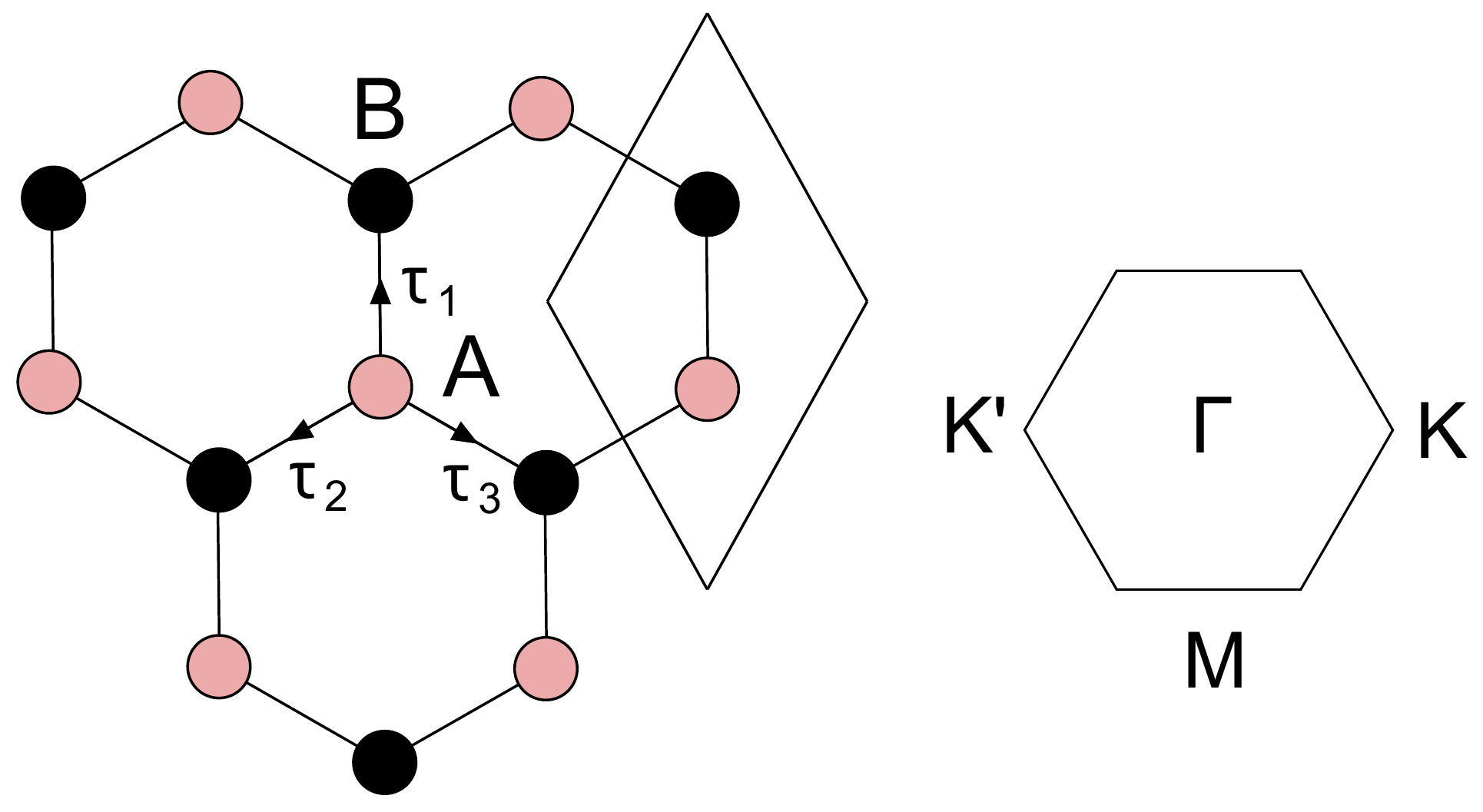} \caption{Left: Honeycomb structure with its two triangular sublattices $A$ and $B$ occupied in hBN by nitrogen and boron atoms, respectively.
$\bm{\tau}_1,\bm{\tau}_2,\bm{\tau}_3$ are the vectors joining first neighbours between the  two sublattices. The vectors are opposite if the origin is taken on a $B$ atom. The unit cell contains one nitrogen and one boron atom. Right: Brillouin zone.}
\label{notations}
\end{center}
\end{figure}

We first specify a few notations. The structure of the hBN single layer is shown in Fig.~\ref{notations}.

\begin{figure}[!ht]
\begin{center}
\includegraphics*[width=7cm]{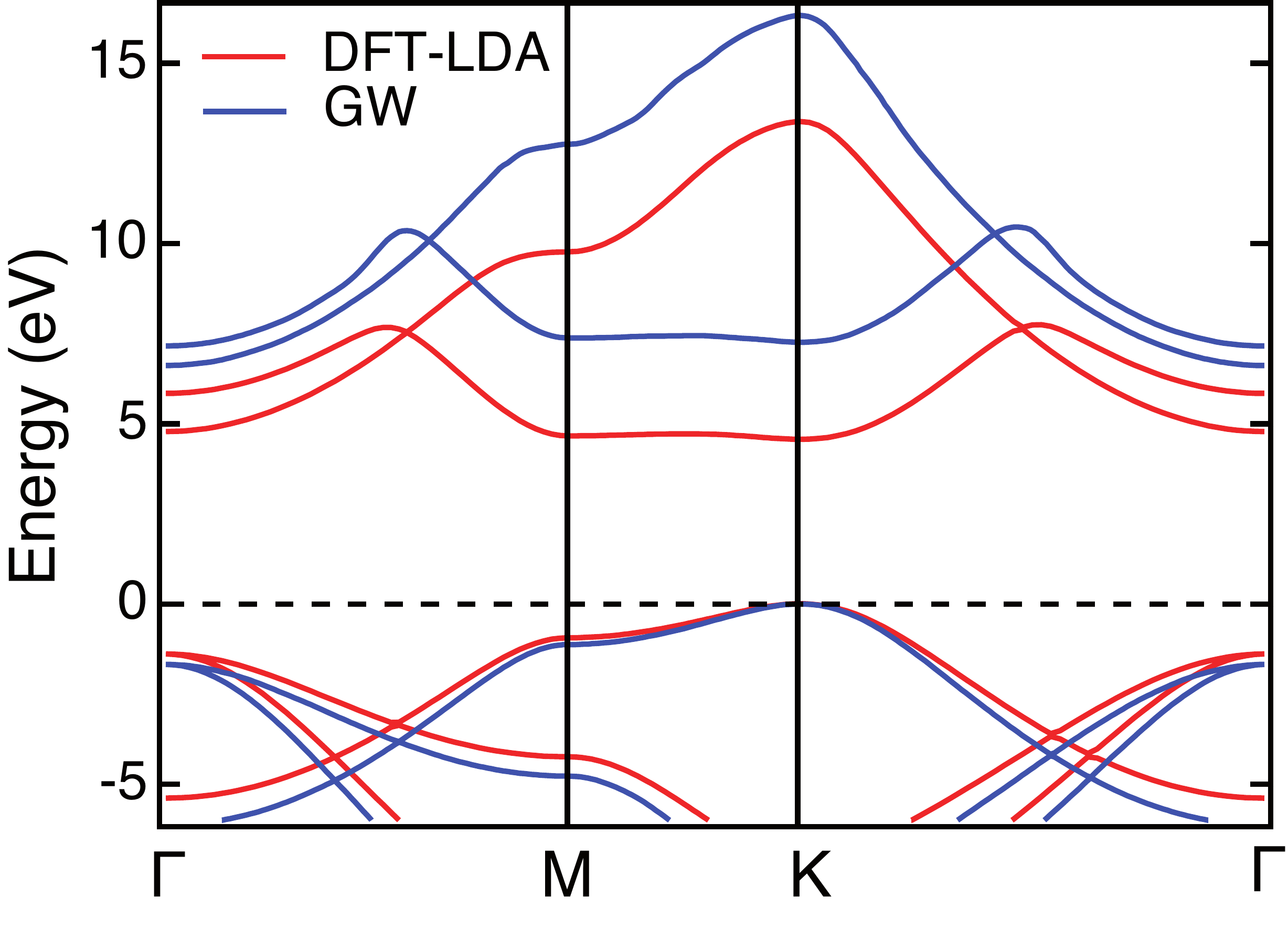} 
\caption{DFT-LDA (red) and GW (blue) \textit{ab initio} band structures of a single hBN layer:  in the $MK$ region, the gap is direct at point $K$ between the flat $\pi$ bands.  The GW corrections were interpolated for this figure with the \texttt{Wannier90} code.\cite{Marzari2008}}
\label{bandabinit}
\end{center}
\end{figure}

The band structure of the hBN single layer is shown in Fig.\ \ref{bandabinit}. The Density Functional Theory (DFT) calculations have been made using the \textsc{Quantum ESPRESSO} code with the local density approximation (LDA) for the exchange-correlation functional. \cite{Baroni2009} The GW corrections were computed in the G$_0$W$_0$ approximation,  using the \texttt{YAMBO} code\cite{Marini2009} with the plasmon-pole approximation for the frequency dependence of the dielectric function. These corrections have been applied to the last two valence bands and to the first four conduction ones. The lattice parameter has been fixed at the experimental lattice constant $a=4.72$ a.u. (2.50~\AA). The computational details are the same used for the subsequent Bethe-Salpeter calculation and are found in Section \ref{section_excitons}.
The gap, equal to 7.25 eV, is direct between the $\pi$ and $\pi^*$ band at point $K$ in the Brillouin zone, while the bands are very flat along the $KM$ lines. This is in agreement with several previous calculations.\cite{Blase1995,Wirtz2009,Ribeiro2011,Berseneva2013,Huser2013,Cudazzo2016} Notice however that the G$_0$W$_0$ approximation is known to underestimate large band gaps. In addition, after the GW corrections are made, the bottom of the conduction band at the $\Gamma$ point is lower than at $K$. This is due to the nearly-free electron states\cite{Blase1995} that are forming around isolated layers of hBN: with increasing inter-layer distance a larger number of states appears. The transition matrix elements from localized valence band states into these states are very low, so for all practical purposes, the isolated sheet of hBN can be considered to be a direct band gap material.
Regardless of the nature of the quasiparticle gap (direct or indirect), the optical gap is direct at point $K$. The \textit{ab initio} results confirm that the contributions to all exciton states of interest in this work come from transitions near $K$ and away from $\Gamma$.

As in the case of graphene, these two $\pi$ bands can be reproduced fairly well using a simple tight-binding model. Let us denote $|\bm{n}\rangle$ the $p_z$ atomic state at site $\bm{n}$. The corresponding atomic orbital is $\phi(\bm{r}-\bm{n}) = \langle \bm{r} | \bm{n} \rangle$. Then, we define the two Bloch functions on the $A$ and $B$ sublattices:
\begin{equation*}
|\bm{k\,} {A(B)}\rangle = \frac{1}{\sqrt{N}}\sum_{\bm{n}\in A(B)} e^{i\bm{K}.\bm{n}}\,|\bm{n}\rangle\; ,
\end{equation*}
where $N$ is the number of unit cells, \textit{i.e.} half the number of atoms. As usual, in most cases we just keep first neighbour hopping integrals $-t, t>0$, and the nitrogen and boron atoms are distinguished by their on-site matrix elements, equal to $-\Delta$ on the $A$ sites for the N atoms, and to $\Delta$ on the $B$ sites for the B atoms. The matrix elements of the hamiltonian in the Bloch basis are therefore written as:
\begin{equation}
\begin{split}
\bra{\bm{k}A}H\ket{\bm{k}A}& = -\Delta \\
\bra{\bm{k}B}H\ket{\bm{k}B}& = +\Delta \\
\bra{\bm{k}A}H\ket{\bm{k}B}& = \bra{\bm{k}B}H\ket{\bm{k}A} ^{*} = -t \, \gamma(\bm{k})\\
\gamma(\bm{k})& = \sum_{\alpha=1,2,3}e^{i\bm{k}.\bm{\tau}_\alpha} \; .
\label{hamilton}
\end{split}
\end{equation}
The energy eigenvalues $E$ are then given by:

\begin{equation*}
E=  s E_{\bm{k}}\quad;\quad E_{\bm{k}}=\sqrt{\Delta^2 +  t^2 |\gamma({\bm{k}})|^2}\quad;\quad s = \mbox{sgn}\,(E) \; ,
\label{gamma}
\end{equation*}
and the eigenstates are:
\begin{equation*}
\ket{\bm{k}\,s} =C_s^A \,\ket{\bm{k}A} + C_s^B \,\ket{\bm{k}B}  \; .
\end{equation*}
Finally, up to a phase factor the coefficients  $C^A_s$ et $C^B_s$ are given by:
\begin{equation}
\begin{split}
C_s^A &= \braket{\bm{k}A}{\bm{k}s} =
- s \frac{\gamma({\bm{k}})}{|\gamma({\bm{k}})|} \sqrt{\frac{E_{\bm{k}}-s\Delta}{2E_{\bm{k}}}}  \\ 
C_s^B &= \braket{\bm{k}B}{\bm{k}s} =\quad
\sqrt{\frac{E_{\bm{k}}+s\Delta}{2E_{\bm{k}}}} \; .
\label{fpropres}
\end{split}
\end{equation}
Thus the $\pi$ electronic structure of hBN-SL can be characterized by only two parameters $t$ and $\Delta$. Their order of magnitude is $t \simeq \Delta \simeq 3$ eV but more precise values can be obtained by fitting the valence and conduction bands $E_{\bm{k}}=\pm\sqrt{\Delta^2 +  t^2 |\gamma({\bm{k}})^2}|$ to those provided by  \textit{ab initio} calculations. $\Delta$ is fixed so that the gap $2\Delta$ is equal to the \textit{ab initio} one, $\Delta = 3.625$ eV, and $t$ is then obtained using standard fitting procedures. Different values are obtained depending on the energy range where the fit is optimized. A global fit, disregarding the nearly-free electron states, leads to $t = 3.0$ eV, but here we are more interested to have a good fit along the MK line, in which case we obtain $t=2.30$ eV (similar to the values for recent fitting of band structures in Ref.~\onlinecite{Ribeiro2011}).

\begin{figure*}[!ht]
\begin{center}
\includegraphics*[width=7cm]{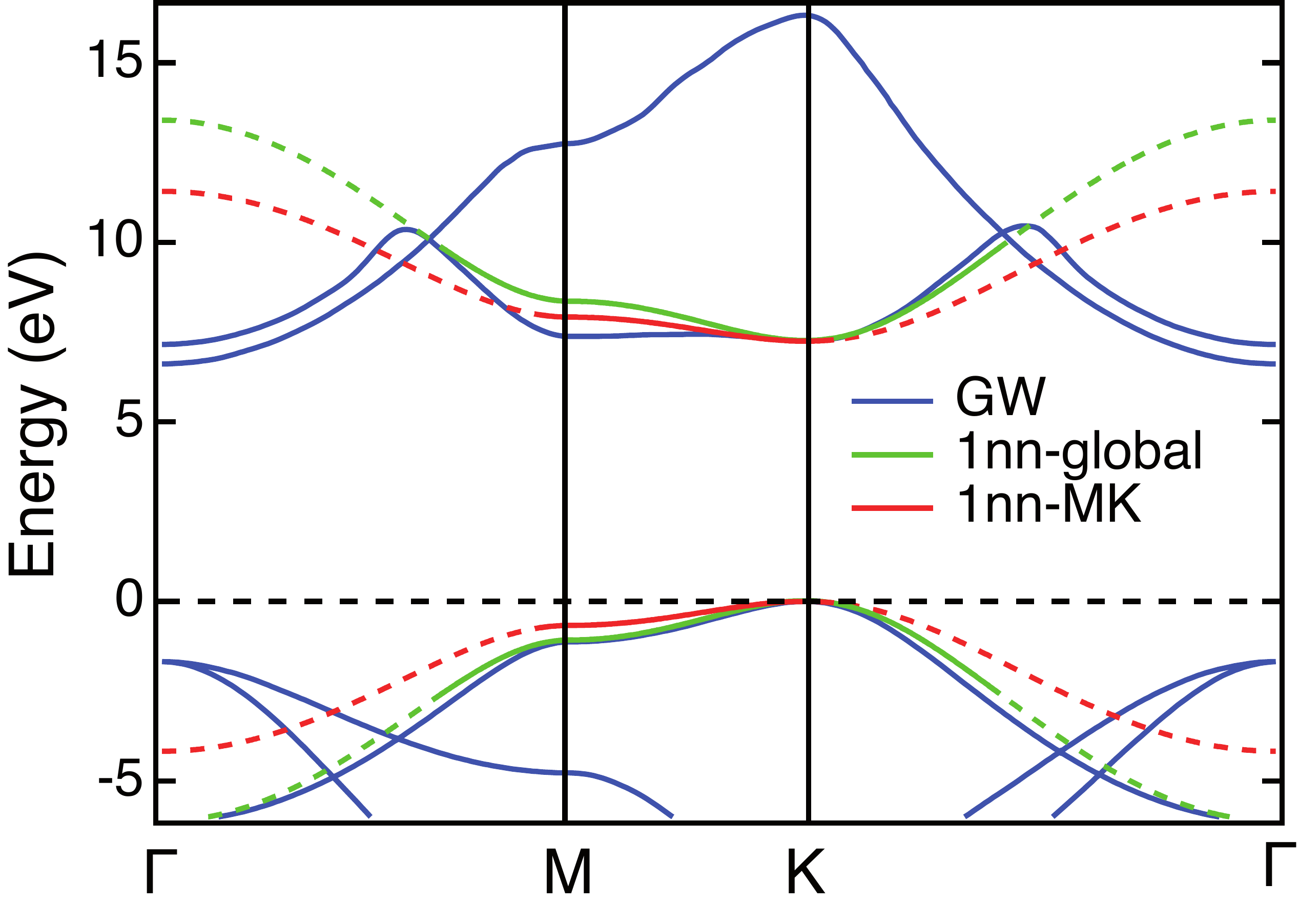}\includegraphics*[width=7cm]{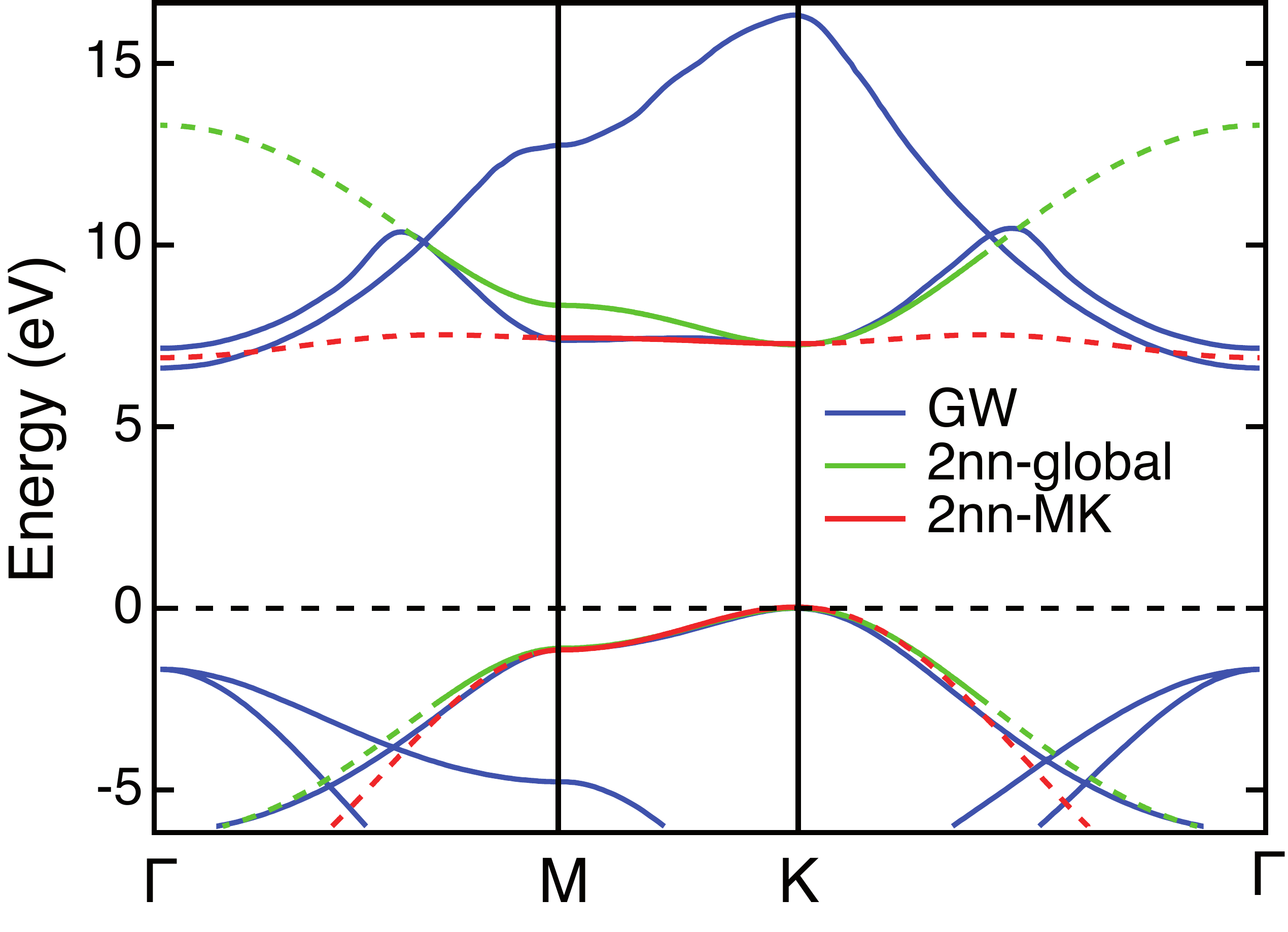}
\caption{Tight-binding fit to the \textit{ab initio} bands including first neighbour interactions (left) and first and second neighbor interactions (right).
Solid lines denote the region of the fit: the global fit is made for all $\pi$ bands except the nearly-free electron states, whereas the $MK$ fits are optimized for the local band structure along the $MK$ lines.}
\label{fitband}
\end{center}
\end{figure*}

As usual, and as shown in Fig.\ \ref{fitband}  the fit is better for the valence band than for the conduction band. The fit can easily be improved by adding further neighbour interactions. Neighbours on the same sublattice contribute to diagonal matrix elements whereas neighbours on different sublattices contribute to off-diagonal elements. 

Before considering  second neighbour interactions explicitly, let us see a simple way to deduce the band structure of hBN-SL from that of graphene. Let $H^0$ be the hamiltonian of graphene with only $t$ hopping integrals. The full hamiltonian $H$  of hBN-SL  is given by  $H= H^0+\hat{\Delta}$ where $\hat{\Delta}$ is the ``atomic'' diagonal hamiltonian with matrix elements equal to $-\Delta$ on the $A$ sublattice and to $+\Delta$ on the $B$ sublattice. It is clear that $H^0\hat{\Delta}+\hat{\Delta} \, H^0 =0$ so that $H^2 = (H^0)^2 + \Delta^2$  where $\Delta^2$ is a simple constant (multiplied by the unit matrix). $(H^0)^2$ on the other hand is an hamiltonian connecting sites entirely on sublattice $A$ or on sublattice $B$. On these triangular lattices $(H^0)^2$ connects the first neighbours with a hopping integral equal to $t^2$ but it has also diagonal on-site matrix elements equal to $3t^2$. In the Bloch basis $(H^0)^2$ is therefore diagonal:
\begin{equation*}
(H^0)^2 = \sum_{\bm{k}}   |\bm{k}A\rangle |t\,\gamma({\bm{k}})|^2 \langle\bm{k}A |+
 |\bm{k}B\rangle |t\,\gamma({\bm{k}})|^2 \langle\bm{k}B| \; ,
\end{equation*}
and the eigenvalues are indeed equal to $\Delta^2 + t^2|\gamma({\bm{k}})|^2$.

Adding second neighbour interactions $-t_2$ in the hBN-SL structure is then equivalent to introducing first neighbour interactions on the triangular sublattices, and the eigenvalues are therefore  given by:
\begin{equation*}
E_{\bm{k}} = -t_2  (|\gamma({\bm{k}})|^2 -3)\,\pm \sqrt{\Delta^2 +  t^2 |\gamma({\bm{k}})|^2} \; .
\end{equation*}
Since both $t$ and $t_2$ are positive, second neighbour interactions induce an asymmetry between the valence and the conduction band: The conduction band becomes flatter than the valence band, in agreement with \textit{ab initio} calculations (Fig.\ \ref{fitband}). The best local fit along $MK$ is provided by $t= 2.30$ eV ; $t_2 = 0.096$ eV. Actually, under the approximation that the valence and conduction bands are pure $N$ and $B$ states, as shown below, only the energy difference between these two bands enters the tight binding excitonic hamilitonian derived in section \ref{section_excitons}, and the second nearest neighbours hopping term does not contribute to this difference. For this reason we limit our tight-binding model for the excitons to first nearest neighbour hopping and keep the simplest previous fit with $\Delta = 3.625$ eV, $ t = 2.30 $ eV.

\subsection{Wave functions, densities of states}

\begin{figure}[!ht]
\includegraphics*[width=8cm]{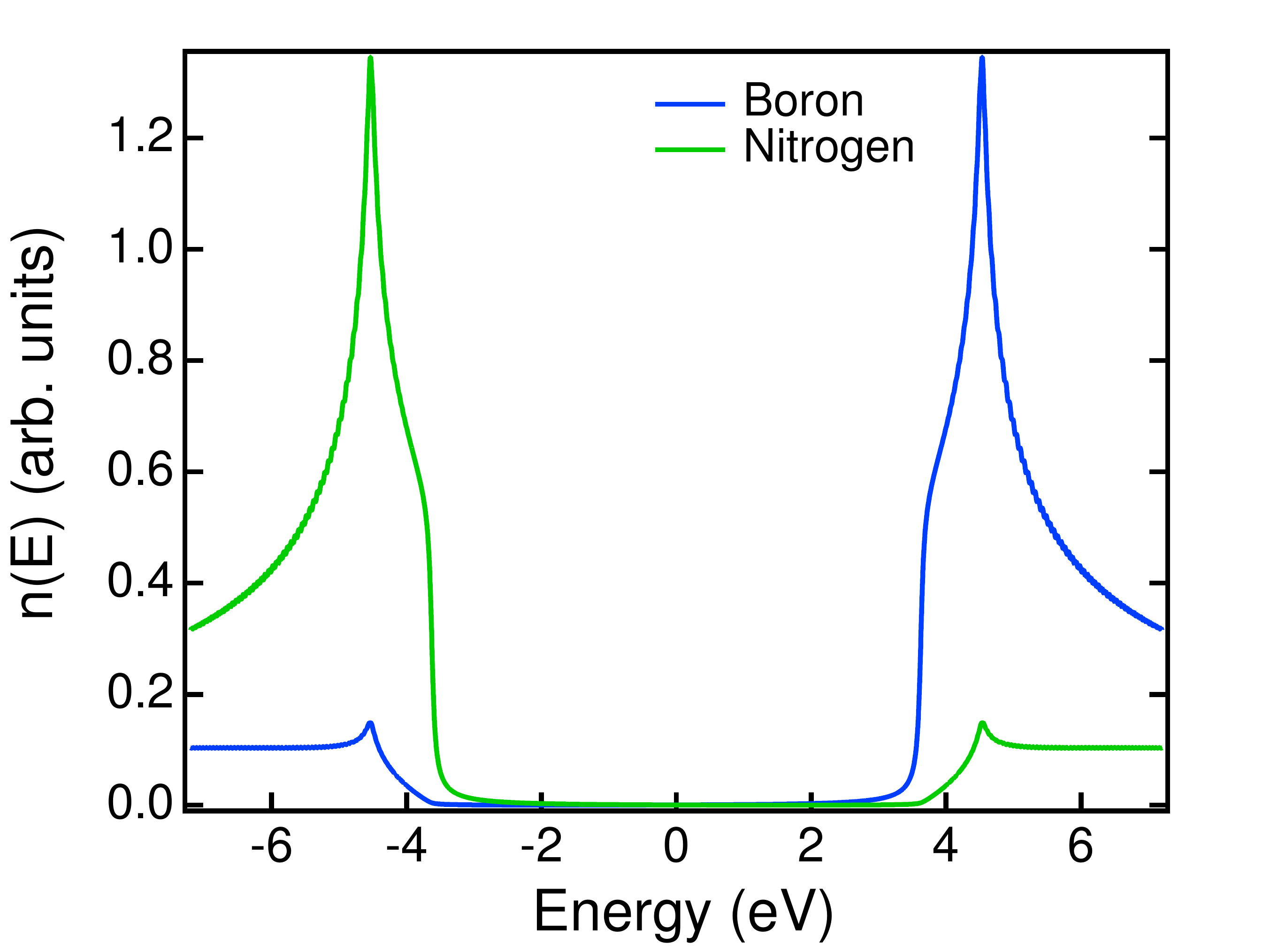} 
\caption{Tight-binding local densities of states on N and B sites calculated using the recursion method ($t=2.30$ eV, $\Delta = 3.625$ eV). The boron (nitrogen) density of states is discontinuous at the upper (lower) band edge $E=\Delta \, (-\Delta)$.}
\label{dos}
\end{figure}

Many electronic properties of hBN-SL only depend on the electronic states close to the gap, \textit{i.e.} in energy ranges where $t |\gamma(\bm{k})|$ is small compared to the gap $2\Delta$. This means that in a first approximation, the coefficients $|C_s^{i}|, i=A,B$ are equal to one or zero. In other words \emph{close to the gap, the valence states are concentrated on the N sites whereas the conduction states are concentrated on the B sites}, and the eigenvalues can be approximated by:
\begin{equation*}
E_{\bm{k}} \simeq   \pm (\Delta + \frac{t^2}{2\Delta} |\gamma({\bm{k}})|^2)  \; .
\end{equation*}
To examine the validity of this approximation, we have calculated the local densities of states $n_{N(B)}(E)$ on both N and B sites.
They are shown in Fig.\ �\ref{dos}. In our simple tight-binding model, one can easily see that $n_{B}(-E) = n_{N}(E)$. Furthermore $n_{B}(E)$ shows a step-function-like onset at $E=+\Delta$, whereas $n_{N}(E)$ has its onset at $E=-\Delta$. As a result the states are indeed quasi-pure B states in a fairly broad energy range above $\Delta$. And of course they are quasi-pure N states below $-\Delta$.

To summarize we can assume that the Wannier functions associated with the valence band and the conduction band can be identified, to lowest order,  with the atomic $\pi$ functions centred on the corresponding sites of their triangular lattices.
The effective hamiltonians $H_v$ for the $\pi$ and $H_c$ for the $\pi^*$ are then given by:

\begin{align}
H_v &= - \sum_{\bm{n}}   |\bm{n}A\rangle \,(\Delta +3t_v)\, \langle\bm{n}A |- \sum_{\bm{n},\bm{m}}{}' \, |\bm{n}A\rangle t_v \langle\bm{m}A| \label{Hv}\\
H_c &= \sum_{\bm{n}}   |\bm{n}B\rangle \, (\Delta +3t_c)\, \langle\bm{n}B |+  \sum_{\bm{n},\bm{m}}{}' \, |\bm{n}B\rangle t_c \langle\bm{m}B| \; ,
\label{Hc}
\end{align}
where the primes indicate sums over nearest neighbours on the triangular lattices, and $t_v=t_c=t^2/2\Delta$.

\section{Excitons in \lowercase{h}BN single layer}
\label{section_excitons}

\textit{Ab initio} excitonic calculations are   based on the Bethe-Salpeter formalism,\cite{Rohlfing2000,Onida2002,Bechstedt2015} which in practice leads to  an effective Schr\"odinger or Wannier equation for electron-hole pairs:
\footnote{Standard treatments of excitons can be found for example in [\onlinecite{Knox1963}],[\onlinecite{Toyozawa2003}], and [\onlinecite{Bechstedt2015}].}

\begin{equation*}
(E_{\bm{k}c} - E_{\bm{k}v})\Phi_{\bm{k}vc} + \sum_{\bm{k'}v'c'} \bra{\bm{k}vc} K_{eh} \ket{\bm{k'}v'c'} \Phi_{\bm{k'}v'c'} = E\  \Phi_{\bm{k}vc} \; ,
\end{equation*}
where $E_{c\bm{k}}$ and   $E_{v\bm{k}}$ are the conduction and valence band energy, respectively, $K_{eh}$ is the electron-hole interaction kernel  and 
$\Phi_{\bm{k}vc}$ is the electron-hole wave function in $\bm{k}$ space. In this paper, we only consider vertical excitations where the electron and the hole have the same wave vector $\bm{k}$, \textit{i.e.} we consider excitons with vanishing wave vector $\bm{Q}$ of their centre of mass. The $\Phi_{\bm{k}vc}$ are  the coefficients in the expansion of the excitonic state $|\Phi\rangle$ in terms of electron-hole excitations:
\begin{equation*}
|\Phi\rangle = \sum_{\bm{k}}\Phi_{\bm{k}vc} \; a^+_{c\bm{k}} a_{v\bm{k}}^{} |\emptyset\rangle \; ,
\end{equation*}
where the vacuum state $|\emptyset\rangle$ is the state where, at zero temperature, all valence states are full and all conduction states are empty. Only singlet states are considered here so that spin indices are omitted. 

The Bethe-Salpeter equation has been solved using the  \texttt{YAMBO} code. \cite{Marini2009}  A Coulomb cutoff of the screened potential in the vertical direction has been  used in order to avoid long-range interaction between repeated copies of the monolayer. \cite{Rozzi2006} In this way, we find that both the GW corrections and the first excitonic peaks are already converged (with about $0.01$ eV accuracy) with an inter-layer separation of $40$ atomic units. The same level of convergence was achieved by sampling the two-dimensional Brillouin zone with a $24\times24\times1$ $k$-point grid. The internal \texttt{YAMBO} parameters for many-body perturbation theory (MBPT) calculations were carefully converged as well. We also carried out calculations with a $36\times36\times1$ $k$-point grid in order to show the higher-level excitonic wavefunctions in real space without any overlap between repeated copies on the same monolayer. We verified furthermore that choosing an inter-layer separation of $80$ a.u. does not modify the results.

\subsection{Ground state exciton}

\begin{figure}[!ht]
\includegraphics*[width=6cm]{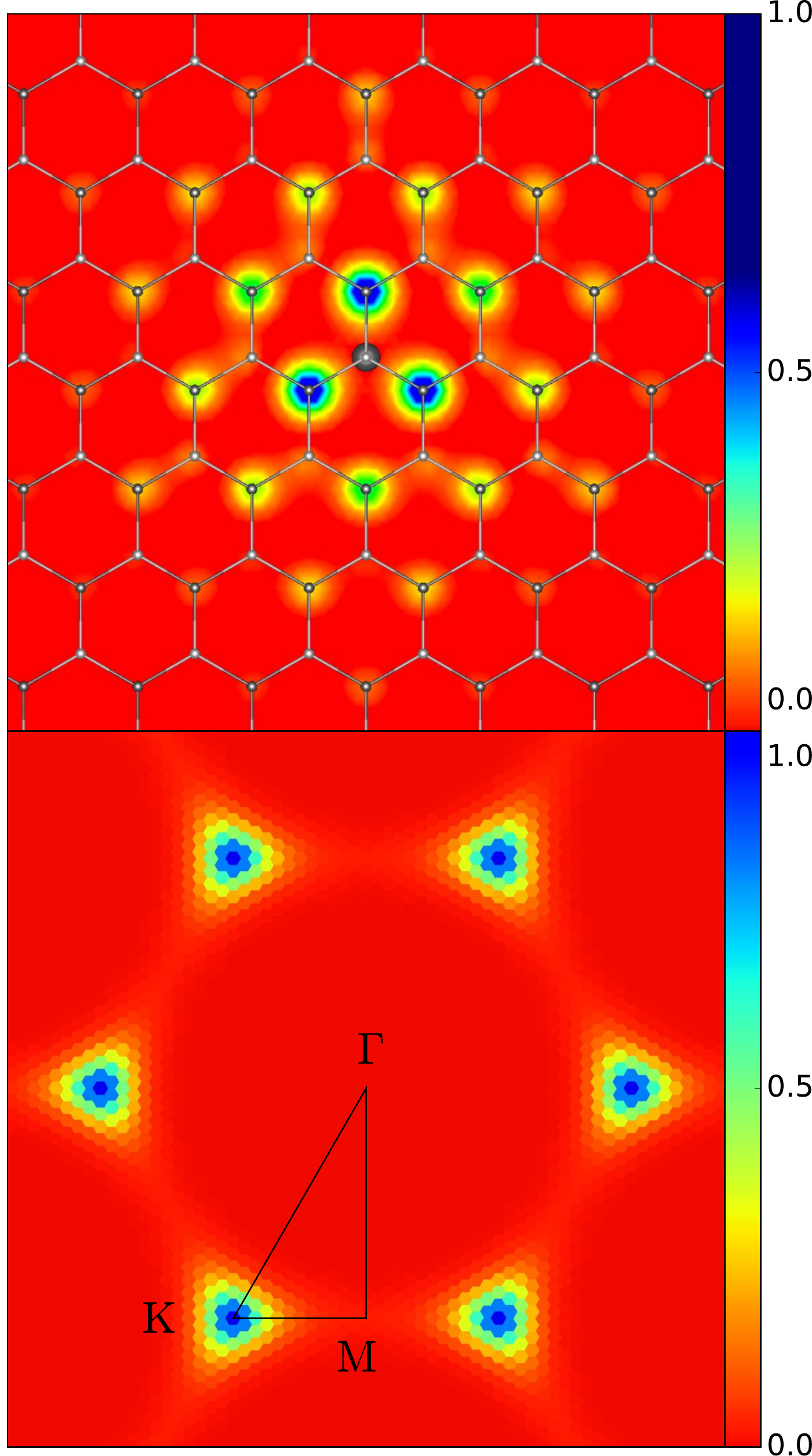} 
\caption {Top: Total probability density $|\Phi(\bm{r}_h,\bm{r}_e)|^2$ of the ground state degenerate exciton states. The hole is located  0.37 \AA \ above the nitrogen atom in the centre (black circle) and $\bm{r}_e$ is at the same altitude. Bottom: Corresponding Fourier intensity $|\Phi_{\bm{k}}|^2$. All intensities are summed over the two degenerate components.}
\label{exciton1}
\end{figure}

As an introduction, we present \textit{ab initio} results concerning the ground state exciton level. Its binding energy measured with respect to the bottom of the conduction band is huge: 1.9 eV. In Fig.\ \ref{exciton1} we show an image of the excitonic wave function $\Phi(\bm{r}_h,\bm{r}_e)$,  where the hole (at $\bm{r_h}$) is localized just above a nitrogen atom. The plot represents the total probability $|\Phi(\bm{r}_h,\bm{r}_e)|^2$, \textit{i.e.} the probability to find the electron at position $\bm{r}_e$ if the hole is located at $\bm{r}_h$. Since this exciton is doubly degenerate, we sum the total probability over the two degenerate states in order to preserve the trigonal symmetry of the crystal lattice.\cite{Wirtz2008}  As expected, the electron density is centred on the boron atoms, with a high probability --- about 30\% --- on the first nearest neighbours. Although not in the genuine Frenkel limit, the exciton is well localized in real space. The shape of this exciton is actually quite similar to that found for the 3D hBN crystal, which is not surprising since in the latter case the exciton was already found to be well confined in a single layer although with a lower binding energy.\cite{Arnaud2006,Wirtz2006,Wirtz2008,Wirtz2009} We also show the wave function in reciprocal space. Here we plot the (summed) weight  $\sum_{vc}|\Phi_{\bm{k}}|^2$ of the electron-hole pairs of wave vector $\bm{k}$ that constitute the bound exciton. The distribution is peaked around the high-symmetry point $K$ but extends toward the boundaries of the Brillouin zone, \textit{i.e.} along the $KM$ lines.%

\subsection{Wannier tight-binding model}

We use now the result derived in Section \ref{section_estructure}  that the valence and conduction states are pure $A$ (N atoms) and $B$ (B atoms) states, respectively. This approximation is fully justified when looking at the \textit{ab initio} results of the previous section. Then we can write:
\begin{equation*}
a_{\bm{k}c}^+ a_{\bm{k}v}^{} \ket{\emptyset} \simeq a_{\bm{k}B}^+ a_{\bm{k}A}^{} \ket{\emptyset}
\simeq \frac{1}{N} \sum_{\bm{n,m}} a^+_{\bm{m}B} a_{\bm{n} A}^{} \; e^{i\bm{k}.(\bm{m-n}}) \ket{\emptyset}.
\end{equation*}
The sum over the electron and hole positions can be decomposed into a sum over the hole position $\bm{n}$ and over the electron-hole distance $\bm{R}$ which is then a vector joining a site on the $A$ (hole) sublattice to a $B$ (electron) sublattice site. For simplicity we use ``bra'' and ``ket'' notations:
\begin{align}
\ket{\bm{k}vc} &= a^+_{\bm{k}c} a_{\bm{k}v}^{} \ket{\emptyset} = \frac{1}{\sqrt{N}}\sum_{\bm{R}} e^{i\bm{k}.\bm{R}} \ket{\bm{R}vc} \\
\ket{\bm{R}vc} &= \frac{1}{\sqrt{N}} \sum_{\bm{n}} a^+_{\bm{n}A+\bm{R}} \; a_{\bm{n} A}^{}  \ket{\emptyset} \; .	
\label{R}
\end{align}
We see that $|\bm{R}vc\rangle$ is the linear superposition of exciton amplitudes for pairs separated by $\bm{R}$. This is nothing but the Bloch wave function for excitons with a wave vector $\bm{Q} =0$. We can then rewrite Bethe-Salpeter-Wannier equation in real space using the $|\bm{R}vc\rangle$ basis and the wave function coefficients $\Phi_{\bm{R}}=  \bra{\bm{R}vc}\Phi\rangle  =  \frac{1}{\sqrt{N}} \sum_{\bm{k}} e^{i\bm{k}.\bm{R}}\; \Phi_{\bm{k}}$. The ``kinetic energy'' term $(E_{\bm{k}c} - E_{\bm{k}v})\Phi_{\bm{k}vc} $, diagonal in $\bm{k}$ space,   becomes a tight-binding-like term in $\bm{R}$ space equal to  $\sum_{\bm{R}'} h_{vc}(\bm{R-R}') \Phi_{\bm{R}'}$ with $h_{vc}(\bm{R}) = \frac{1}{N} \sum_{\bm{k}} e^{i\bm{k}.\bm{R}} \; (E_{c\bm{k}} - E_{v\bm{k}})$. 
To be more precise let $H_{vc}$ be the hamiltonian acting in the excitonic space. The kinetic energy (or free single-particle) term is the difference of the two 
hamiltonians \eqref{Hv} and \eqref{Hc}.  We rewrite it here $H^0_{vc} =  H_c\otimes 1 -1\otimes H_v$ to indicate that each term in the r.h.s. acts either on the electronic or on the hole component of the electron-hole states. Dropping now the $v,c$ indices within the ``bras'' and ``kets'', the matrix elements $\langle\bm{R}|H_{vc}|\bm{R}'\rangle$ of $H_{vc}^0$ are  equal to $h_{vc}(\bm{R}-\bm{R}') $ and are therefore equal to $2\Delta + 3t^2/\Delta$ if $\bm{R}=\bm{R}'$, and to $t^2/\Delta$ if $\bm{R}$ and $\bm{R}'$ are first neighbours on the triangular lattice.

The electron-hole interaction kernel on the other hand contains a direct  term and an exchange contribution. Let us first consider the direct term which is the most important one, as will be checked later (see Table \ref{tabexi}). The corresponding integral can be expanded in real space and involve integrals of type:
\begin{widetext}
\begin{equation*}
-\int d\bm{r} d\bm{r}' \varphi_v(\bm{r}-\bm{R}_n)   \varphi_v(\bm{r}-\bm{R}_p)  \frac{e^2}{|\bm{r}-\bm{r}'|} 
 \varphi_c(\bm{r}'-\bm{R}_m)  \varphi_c(\bm{r}'-\bm{R}_q)   \; ,
\end{equation*}
\end{widetext}
where the $\varphi_{v(c)}(\bm{r})$ are (real) valence or conduction orbitals.

Usually the largest integral is the one where all indices are identical, but here this on-site integral is forbidden since the conduction and valence orbitals belong to different sublattices (actually, this is not completely true as will be discussed below in Section \ref{wannier}).  The next most important integrals are those where $p=n$ and $q=m$, and finally the (direct) Coulomb term $\sum_{\bm{k'}} \bra{\bm{k}vc} K_{eh}^d \ket{\bm{k'}vc} \Phi_{\bm{k'}vc}$  becomes $\sum_{\bm{R}\neq 0 } U_{\bm{R}} \Phi_{\bm{R}}$ where:
\begin{equation*}
U_{\bm{R}} = \bra{\bm{R}} K_{eh}^d \ket{\bm{R}} =
 - \int d\bm{r} d\bm{r}' \varphi_c^2(\bm{r})\frac{e^2}{|\bm{r-r}'|} \varphi_v^2(\bm{r}'-\bm{R}) \; ,
\end{equation*}
which means that $U_{\bm{R}}$ acts as a local potential on ``site'' $\bm{R}$. The Coulomb potential should also be screened but here the $U_{\bm{R}}$ will just be considered as parameters to be fitted to \textit{ab initio} data.

We have therefore reduced our problem to a very simple tight-binding problem for the relative motion of the electron and of the hole. Since the motion is relative we can fix the hole at the origin of the $A$ sublattice. The  $\bm{R}$ vectors lie on the $B$ sublattice: our problem becomes the problem of an electron moving on the $B$ sublattice in the presence of a hole at the origin which plays the part of an impurity, source of the attractive potential $U_{\bm{R}}$. To summarize, when exchange effects are neglected, we have to handle the standard tight-binding equations:
\begin{equation*}
E \Phi_{\bm{R}} = \sum_{\bm{R}'} h_{eh}(\bm{R-R}') \Phi_{\bm{R}'} + \sum_{\bm{R}} U_{\bm{R}} \Phi_{\bm{R}}   \; ,
\end{equation*}
which therefore depends only on $t_{ex}=t^2/\Delta$ and on $U_{\bm{R}}$.

\subsection{Discussion of the Wannier model}
\label{wannier}

Although standard, the Wannier equations are difficult to solve  in many cases because several valence and conduction bands are involved. As a consequence the Wannier functions have no longer direct relationships with the atomic orbitals. On the other hand, in the case of strong screening and small gaps, the potential does not perturb the single particle Bloch states too much and $\bm{k}.\bm{p}$ expansions can be used, leading to the familiar hydrogenic model where the underlying lattice can finally be forgotten. This model is \textit{a posteriori}  justified when the extension of the excitonic states is large compared to the lattice parameter. As shown in Fig.\ \ref{exciton1} this is not  the case here, but the simplicity of the electronic structure of the boron nitride single sheet will allow us to take lattice effects fully in account. 

\subsubsection{A very simple model}

\begin{figure}[!ht]
\includegraphics*[width=5cm]{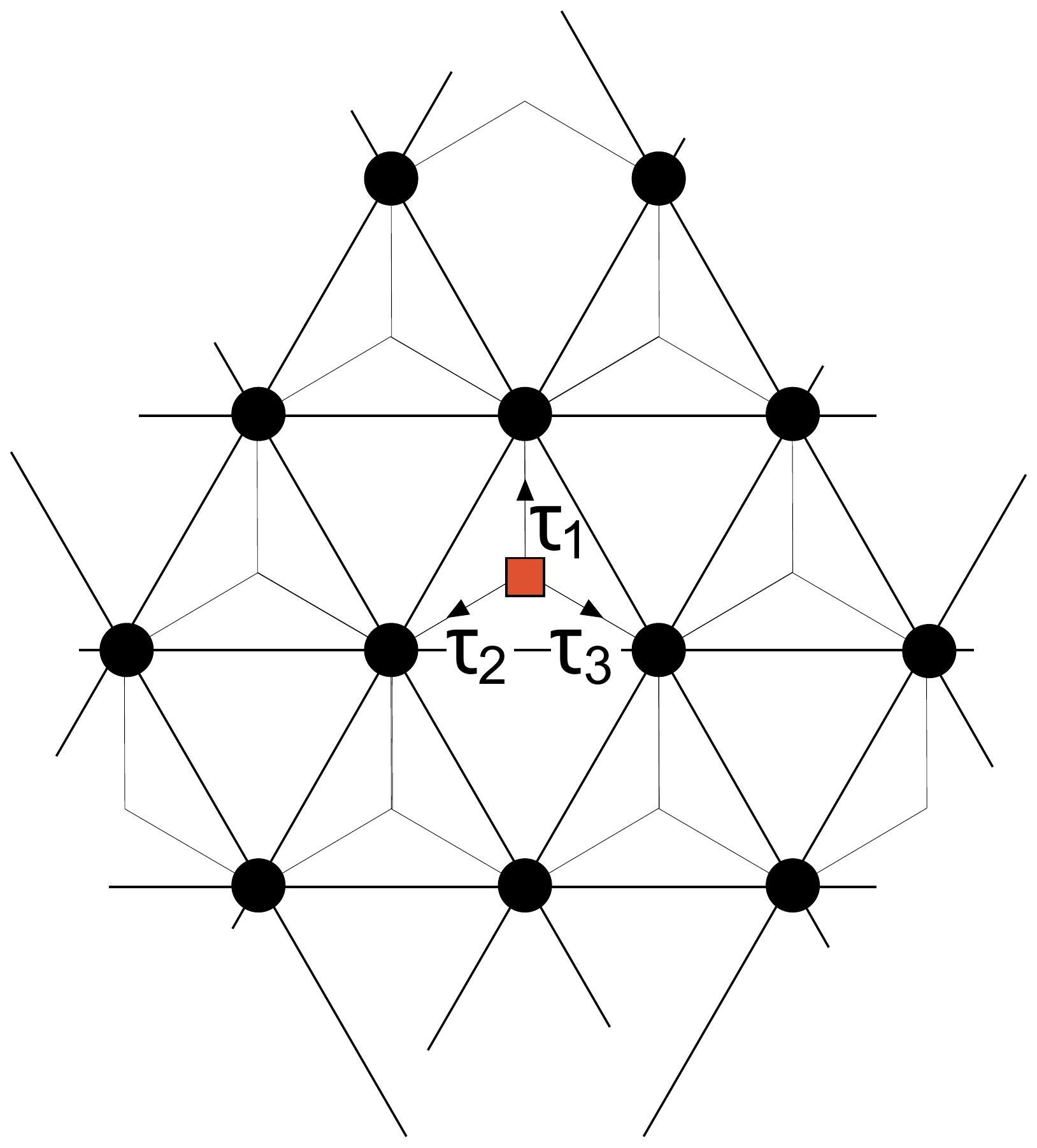} \quad\quad
\includegraphics*[width=8cm]{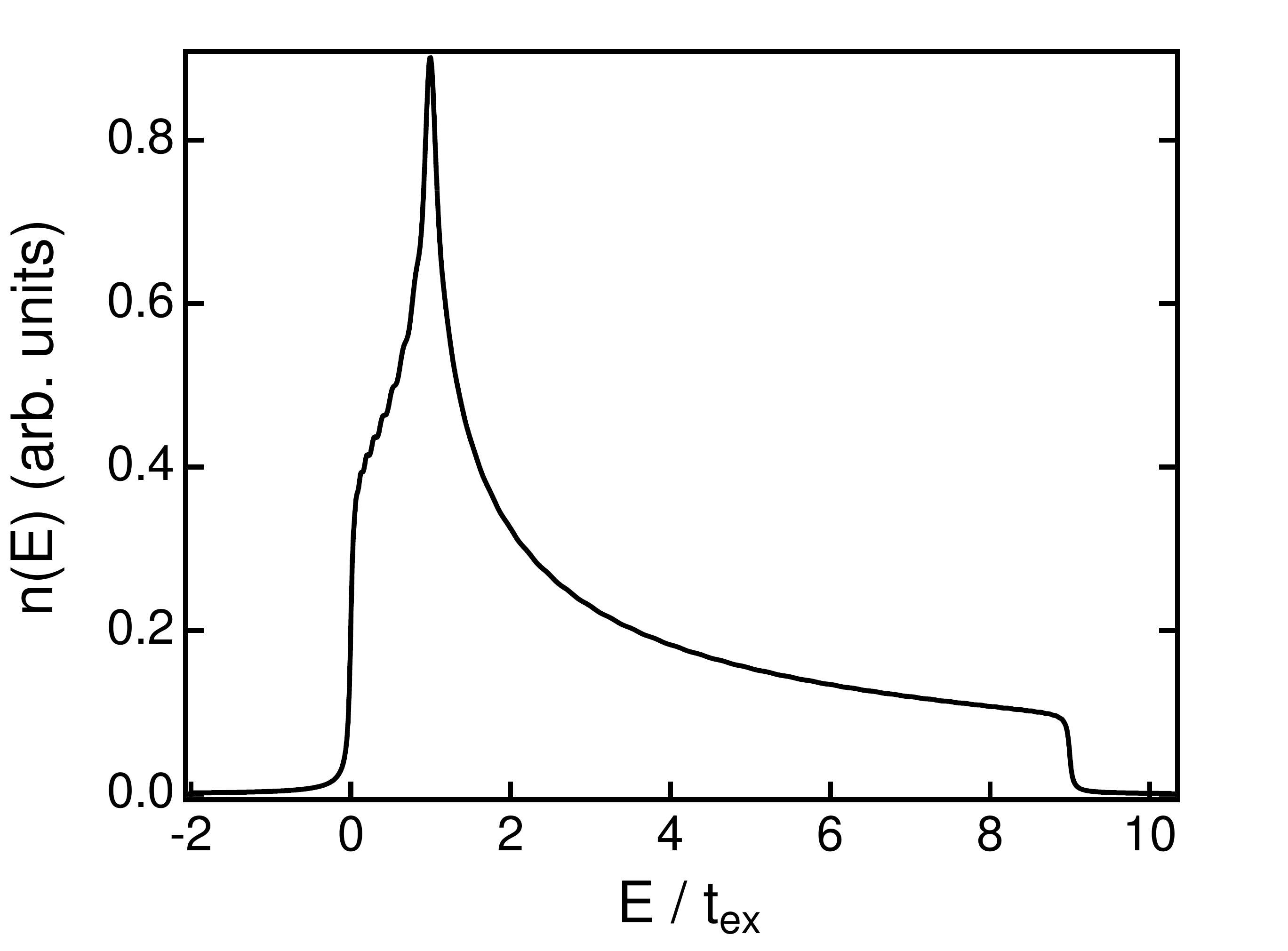} 
\caption{Left: The hole is at the origin (square) and the electron is moving on a triangular sublattice. Right: Density of states corresponding to the spectrum of the excitonic hamiltonian without Coulomb interactions, $H^0_{eh} -2\Delta$, so that the spectrum starts at $E=0$. The Van Hove singularity is at $E=t_{ex}=1.46$ eV.}
\label{triangle}
\end{figure}

As shown above, we have to solve an impurity problem in a simple tight-binding basis. In the case of localized potentials, Green function or direct diagonalization techniques are known to be very efficient. Actually we have just to transpose the methods used to study deep impurities centres in semiconductors.\cite{Lannoo1981,Yu2010} Although the Coulomb potential is a long range $1/R$ potential we expect that the energy of the lowest bound state can reasonably be obtained using truncated potentials in real space. Let us recall that the electron is moving on a triangular lattice with first neighbour hopping integrals in the presence of an impurity located at the origin taken at a lattice point of the hole sublattice, \textit{i.e.}  at the centre of a triangle of the B sublattice (see Fig.\ \ref{triangle}). We have therefore to diagonalize the following hamiltonian:

\begin{align*}
H_{eh} &= H^0_{eh} + U \\
\bra{\bm{R}}H^0_{eh}\ket{\bm{R'}} &= 2\Delta + 3t^2/\Delta\quad \mbox{if}\quad  \bm{R} =\bm{R}' \\
&= t^2/\Delta \quad \mbox{if} \; \bm{R} \; \mbox{and} \; \bm{R}' \; \mbox{are first neighbours}\\
& =0\quad  \mbox{otherwise} \\
U &= \sum_{\bm{R}} \ket{\bm{R}} U_{\bm{R}} \bra{\bm{R}} \; ,
\end{align*}
where the potential is to be fitted to \textit{ab initio} data, and where exchange terms are neglected.
The spectrum of $H^0_{eh}$ is known since it is the spectrum of the triangular lattice with (positive) first neighbour hopping integrals (Fig.\ \ref{triangle}). The lowest eigenvalue (taken as the origin in Fig.\ \ref{triangle}) is at $2\Delta$, which corresponds to the energy gap in this model. In the presence of a localized attractive potential, we expect that bound excitonic states appear when the potential is strong enough. This is indeed what happens. A brief analytical discussion is presented in Appendix \ref{greenf}. Here, we present results  obtained with a potential fitted up to  the 28th neighbouring shell to \textit{ab initio} data, and which is discussed below, in Sec. \ref{fit}. We have diagonalized our tight-binding hamiltonian using a box containing about $10^3$ sites on the triangular lattice. Many exciton states have been studied, but since our model is obviously becoming inaccurate when the energy rises, and when the extension of the exciton increases, five states are just  considered in detail.  The advantage of our procedure in real space is that we handle real wave functions and so we can simply image the wave function themselves. Furthermore in the case of degenerate states it is easy to show components of definite symmetry.

\subsubsection{The ground state exciton and the exciton symmetries}

\begin{figure}[!ht]
\includegraphics*[width=8.5cm]{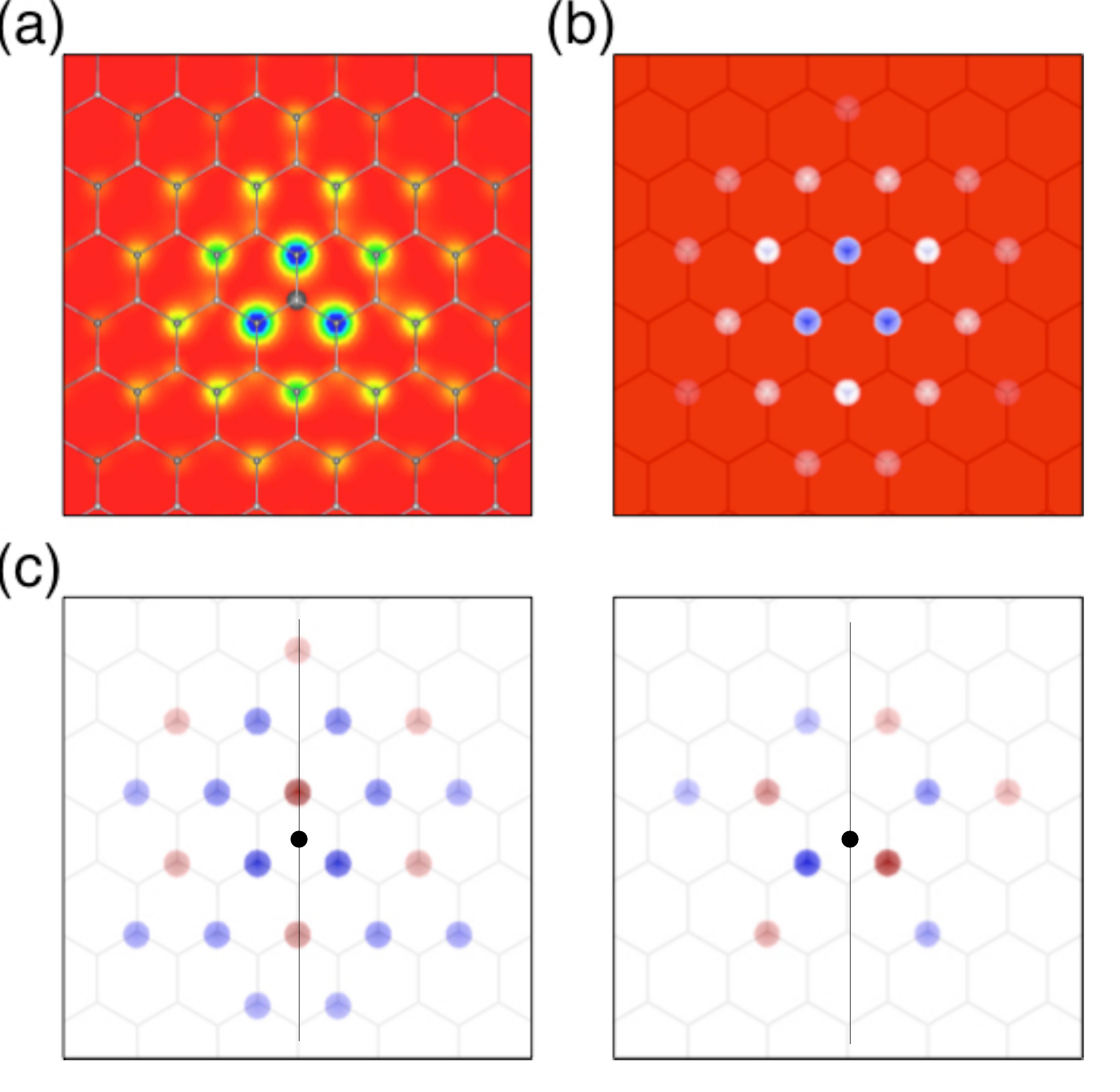} 
\caption{Results for the ground state exciton. a) \textit{Ab initio} intensity.  b) Tight-binding intensity, and c) tight-binding amplitudes for the two degenerate states, symmetric and antisymmetric  with respect to the $y$-axis. Blue and red colours in c) correspond to opposite signs.}
\label{exc1ter}
\end{figure}

We show first  in Fig.\ \ref{exc1ter} the results concerning the ground state exciton which is doubly degenerate. In the tight-binding case, we show the two components which are clearly antisymmetric or symmetric with respect to the $y$-axis. The agreement with the \textit{ab initio} result is very good. At this point it is useful to comment on the symmetry of this state. Since we have fixed the position of the hole, we can use the point symmetry of the triangular lattice with respect to the origin located at a centre of a triangle. In principle the problem is not purely a 2D one since the $\pi$ orbitals extend in the $0z$ direction and are odd with respect to a  $z \to -z$ reflection. Apart from this trivial symmetry, we have just to consider the $C_{3v}$ symmetry with its 3-fold rotation axis and its mirror planes $\sigma_v$. This group is known to have three different representations. Beyond the identity one, $A_1$ we have the familiar 2-dimensional representation $E(x,y)$, and a second representation of dimension 1, $A_2$ characterized by an odd character for the $\sigma_v$ reflections. Our exciton has clearly here the $E$ symmetry with two (chiral) components which can be taken as varying as $x+iy$ or $x-iy$.

This exciton is very similar to the so-called $A$ or $B$ excitons met in TMD. In this case, the Wannier-Mott approach in the $\bm{k}.\bm{p}$ approximation is generally used, and the symmetry of the excitons is frequently defined as follows: The exciton wave function is written in the form $\Phi(\bm{r}_h,\bm{r}_e)=\phi_{\bm{k_0}c}(\bm{r}_e)\phi_{\bm{k_0}h}(\bm{r}_v)g(\bm{r}_e-\bm{r}_h)$, where the $\phi_{k_0}$ are the single particle Bloch functions at point $\bm{k_0}$ corresponding to the considered direct gap, and $g(\bm{r})$, the envelope function,  is the solution of the hydrogenic-like excitonic equation for the relative coordinate $\bm{r}=\bm{r}_e-\bm{r}_h$.\cite{Jorio2011} The full excitonic symmetry is the symmetry of this product, but notations generally use the symmetry of $g(\bm{r})$. This decomposition makes sense uniquely if the $\bm{k}.\bm{p}$ expansion around $\bm{k}_0$ is valid, which is not necessarily the case here. In our case, the direct gap occurs at points $K$ and $K'$. Neglecting intervalley coupling (which may not be valid either), we see that  $\bm{k}_0=\bm{K}$ and the product of Bloch functions $\phi_{\bm{k_0}c}(\bm{r}_e)\phi_{\bm{k_0}h}(\bm{r}_v)$ varies as $e^{i\bm{K}.\bm{r}}$, \textit{i.e.} as one $E$  component of the representation of the $C_3$ symmetry at point $K$. Now, since the conduction and valence bands are non degenerate at point $K$, the ground state envelope function $g(\bm{r})$ is isotropic and of symmetry $s$. This is why the $A$ exciton is denoted a $1s$ exciton. With similar arguments we obtain that the exciton at $K'$ has also a $s$ symmetry modulated by the Bloch function proportional to $e^{-i\bm{K}.\bm{r}}$. So, in this description we obtain two degenerate $1s$ excitons, but if they are considered together they form a double degenerate exciton of symmetry $E$. Both descriptions are equivalent in the case of large excitons which can be associated separately to points $K$ and $K'$.\cite{Wu2015} In our case where the exciton is localized in real space (delocalized in reciprocal space), using directly the full point symmetry of the exciton is more accurate.

\subsubsection{Other excitons}

\paragraph{Analysis in real space}

\begin{figure*}[!ht]
\includegraphics*[width=17cm]{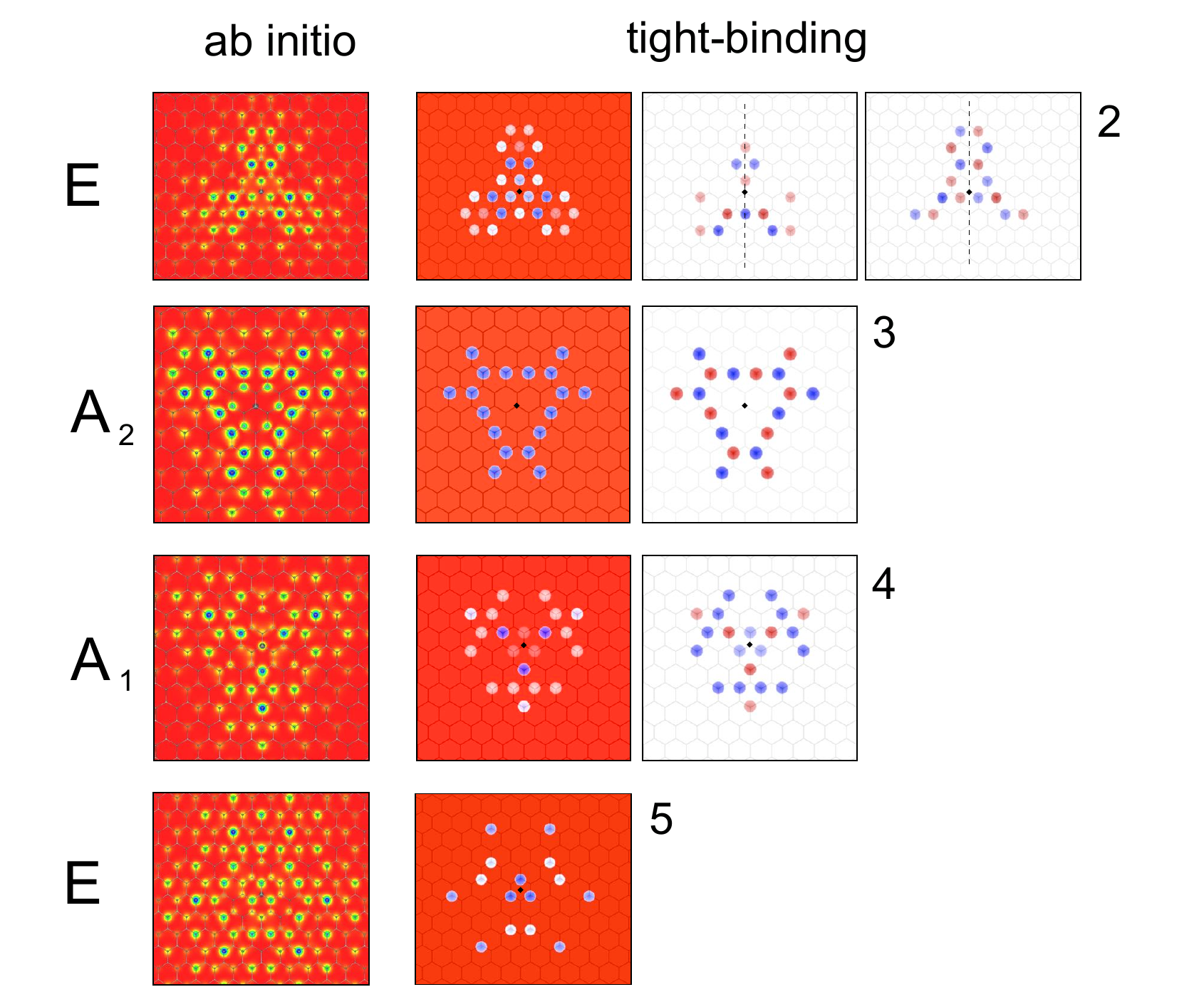} 
\caption{Results for the  excitons \#2 to \#5. Left: \textit{ab initio} intensity. Right: tight-binding intensities and amplitudes. The hole is at the centre of the central triangle. Blue and red colours in these plots correspond to opposite signs. Exciton \#2 has two components as the ground state exciton. The tight-binding analysis shows clearly that exciton \#3 has the $A_2$ symmetry with an anti-symmetric behaviour with respect to the three $\sigma_v$ mirrors. Notice in the \textit{ab initio} image the low intensity in the interior hexagon, \textit{i.e.} on N sites. This is a signal that the simple tight-binding model which forbids this possibility begins to fail. But otherwise, the agreement is very good. Notice also that, due to the $A_2$ symmetry, the intensity strictly vanishes on the first neighbours and more generally on the symmetry axis. Exciton \#4 on the other hand as the full $C_{3v}$ symmetry typical of the identity representation $A_1$. The \textit{ab initio} results show a significant intensity on the central N site. Finally exciton \#5 has a $E$ symmetry. Since the amplitude images are fairly complex only the tight-binding intensity is shown.}
\label{exc2-5}
\end{figure*}

At higher energy, a group of six states appears. All of them as well as the previous state have  similar energies within 0.1 eV. Actually they do not appear in the same order in both \textit{ab initio} and tight-binding calculations. Their wave functions are on the other hand very similar. We follow here the order provided by the \textit{ab initio} calculations. Both \textit{ab initio} and tight-binding approaches find first a similar exciton with again a twofold degeneracy (exciton \#2 in Fig.\ \ref{exc2-5}). It has therefore also an $E$ symmetry. The agreement between both calculations is still fairly good. The two following ones are non degenerate. The TB method shows unambiguously that the first one  transforms according to the $A_2$ representation (exciton \#3, Fig.\ \ref{exc2-5}) and the second one according to the  $A_1$ one (exciton \#4, Fig.\ \ref{exc2-5}). Finally, the two upper states are degenerate and belong to the $E$ symmetry (fifth exciton in Table~\ref{tabexi}). We summarize in Table \ref{tabexi} the energies and symmetries of these excitons. The next  excitons are found more than  0.2 eV above this group in the \textit{ab initio} calculations.

\begin{table*}[!ht]
\centering
\begin{tabular}{|c|c|c|c|c|c|c|}
\hline
Exciton	&1 (x2)	&2 (x2)	&3	&4	&5 (x2) \\ \hline
Ab initio  &-1.932& -1.076	&-1.045	&-0.980	&-0.892			\\ \hline
Ab initio without exchange		&-2.018&-1.095&-1.045&-1.358&-0.898	\\ \hline
Tight binding  &-1.932 & -1.053	&-0.999	&-1.0944	&-0.830		\\ \hline
Symmetry  &$E$	 & $E$	&$A_2$	&$A_1$	&$E$	\\ \hline	
\end{tabular}
\caption{The five first excitons in the order fixed by the \textit{ab initio} calculations.
Energies are in eV.}
\label{tabexi}
\end{table*}

Although the overall agreement between \textit{ab initio} and TB calculations is fairly good, the behaviour of the $A_1$ exciton seems particular. This is still more obvious if we compare \textit{ab initio} calculations performed with and without the exchange contribution (Table \ref{tabexi}). Whereas the energy variation between the two calculations for the other excitons is of a few percents, the $A_1$ exciton is strongly perturbed, its energy moving from $-0.98$ eV to $-1.358$ eV when the (repulsive) exchange contribution is suppressed. This is quite unusual but can be related to the fact that the $A_1$ exciton is, by symmetry,  the only one where the electronic intensity at the origin is non-vanishing. Actually, although we have neglected this possibility in our simplified TB model, the \textit{ab initio} calculations do show such a non-vanishing intensity (Fig.\ \ref{exc2-5}). The point is that even if this intensity is low, it introduces a perturbation proportional to the on-site exchange term $2 J_0$, which is very large. Actually, the local Coulomb and exchange integrals $U_0$ and $J_0$ are of the same order of magnitude since they both involve similar $\pi$ orbitals. They are of opposite signs however, which explains why our TB scheme which neglects on-site Coulomb interactions is not too bad even in this case. \\

\paragraph{$2s$ and $2p$ states: Analysis in reciprocal space}

\begin{figure}[!ht]
\includegraphics*[width=8cm]{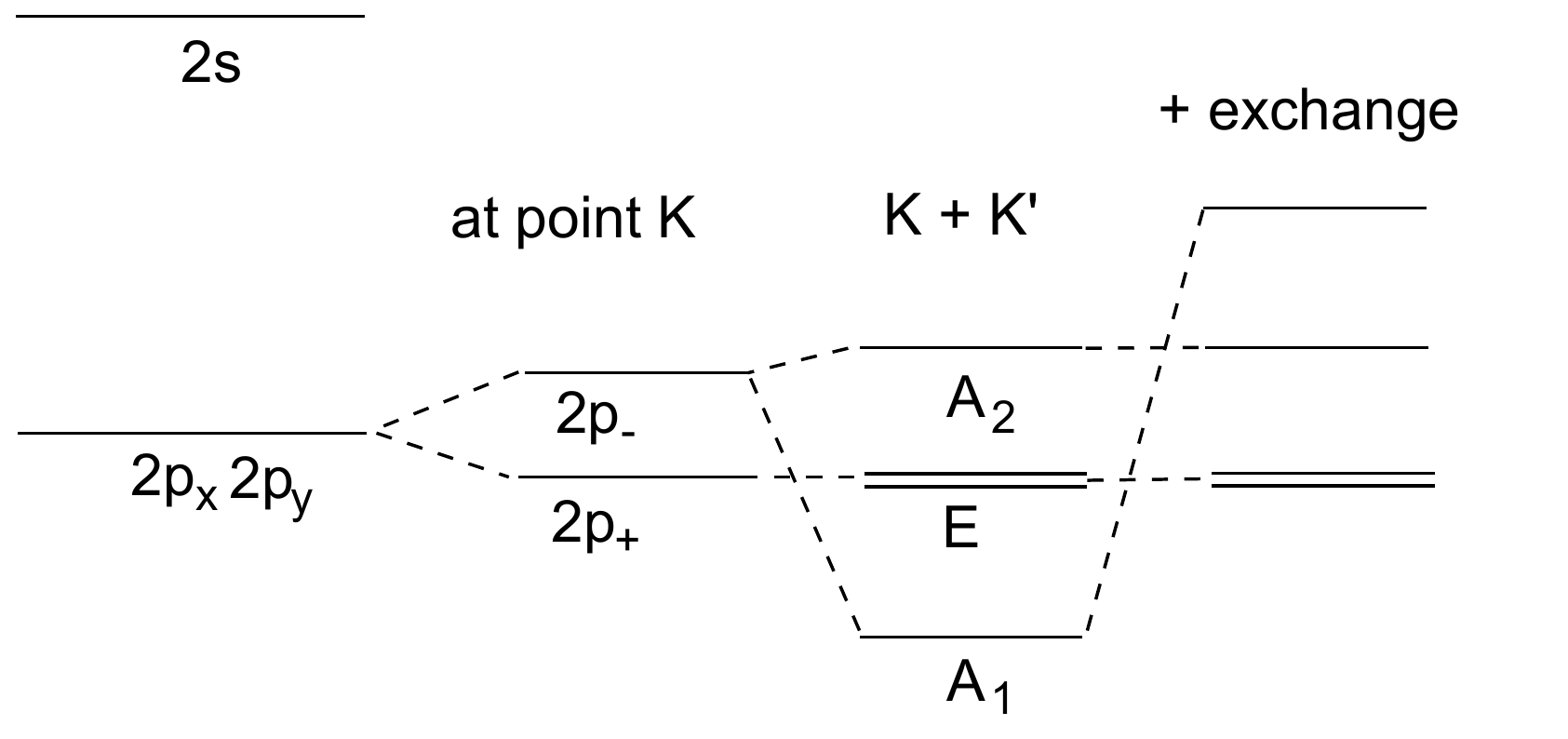} 
\caption{Schematic splitting scheme of the $2p$ levels. Depending on the calculations, \textit{ab initio} or TB, the level separations are of the order of 0.1 eV or less. The $1s$ state is about 1 eV below. }
\label{2p_level}
\end{figure}

At this point it is  instructive to compare our results with those obtained for dichalcogenides. In TMD the Wannier-Mott model is valid provided appropriate anisotropic potentials are used. It is then convenient, as discussed earlier, to analyse the excitons in each valley in terms of $s, p, \dots $ symmetries. The usual sequence is a $1s$ level, and then a $2s$ level nearly degenerate with a $2p$ level. This $2p$ level gives rise to four states because of the valley degeneracy. A careful examination of the symmetry of the $\pi$ and $\pi^*$ states close to the $K$ points based on the so-called massive Dirac model has shown that actually the degeneracy within each valley is lifted, but  time reversal symmetry between $K$ and $K'$ insures that the $2p$ states are splitted into two doubly degenerate states. Furthermore the $2s$ level is found to be above the $2p$ levels.
\cite{Wu2015,Srivastava2015,Zhou2015,Berkelbach2015} In our case where lattice effects and therefore inter-valley effects are included a further splitting occurs. As pointed out above, the full symmetry of the exciton states is obtained from the product of the envelope function ($E$ symmetry for $p$ states) and of the Bloch functions at points $K$ of symmetry $E$ also. Now, the decomposition of the $E\times E$ representation gives rise precisely to the observed one: $E\times E = E+A_1+A_2$. In the $s,p, \dots$ language, the $p$ states are first splitted into $p_x\pm ip_y$ states whose chiralities are equal or opposite to those of the Bloch functions at points $K$ and $K'$. Hence two states (one in each valley) have a vanishing global chirality. Forming bonding and antibonding states between these states lead to the $A_1$ and $A_2$ states. The two other ones remain degenerate and form a $E$ state. We expect the  bonding state  $A_1$ to be below $A_2$, but, as argued before, the repulsive exchange contribution neglected in this discussion pushes the $A_1$ upwards. This is described in Fig.\ \ref{2p_level}. 

\begin{figure*}[!ht]
\includegraphics*[width=12cm]{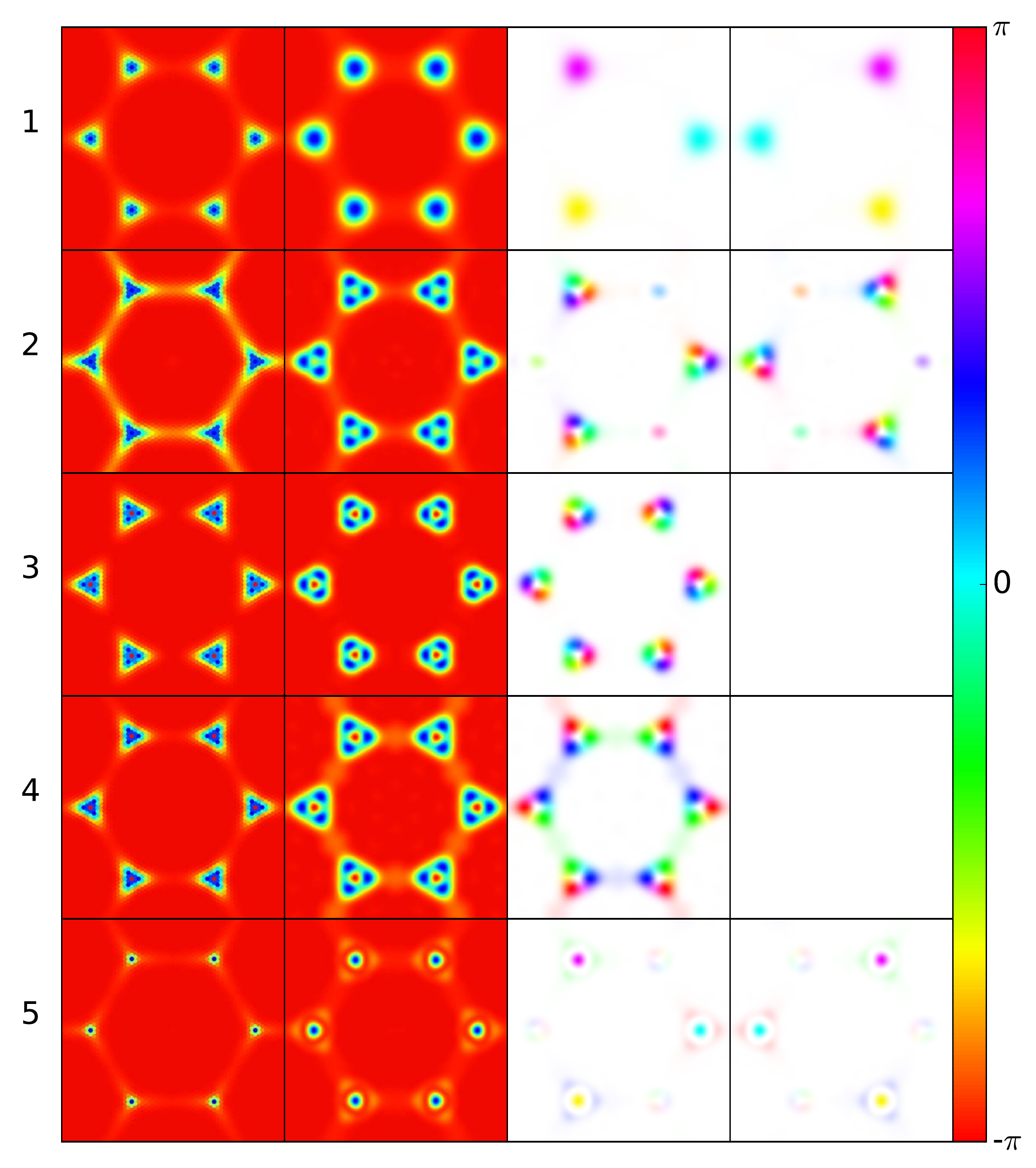} 
\caption{\textit{Ab initio} and TB results for the intensities and wave functions in reciprocal space for the excitons \#1 to \#5. First column: \textit{Ab initio} results. The first and last excitons show intensities peaked at $K$ and $K'$, more concentrated for the last $2s$ state than for the first $1s$ one, as expected. This is actually the way state \#5 is identified as $2s$. The three other states (four, including degeneracy) have similar shapes with significant trigonal warping effects. The TB results are shown in the second column and are very similar. The last two columns show the TB wave function with a modulus-phase representation: a circle with an opacity proportional to the intensity at this $k$-point is placed on each point of a grid. The colour of the circle is related to the phase between $-\pi$ and $+\pi$ as indicated in the colour bar. Two plots are shown in case of degenerate states which have been filtered according to the chiral symmetries. It is clear that the phases are equal to $0$ and $\pm 2\pi/3$ at these points for the $1s$ and $2s$ states. The four other states have more rich structures with phases rotating within each triangular spot located at the K points and, as explained in the main text, are a signature of ``$2p$'' states.} 
\label{recip}
\end{figure*}

It remains to determine which $E$ exciton belongs to this $2p$ family. This might be exciton \#2, as shown in the figure, or exciton \#5. It is not obvious to decide from the plots of the wave functions in real space, but we  show now that an analysis in reciprocal space provides the answer. In the first column of  Fig.\ \ref{recip}  are shown the intensities $|\Phi_{\bm{k}}|^2$ of the five excitons considered previously. It is clear at once that excitons \#2, \#3, and \#4 belong to the same family and are therefore the expected $2p$ excitons, which means that the relevant $E$ exciton is exciton \#2 as depicted in Fig.\ \ref{2p_level}. A consequence is that exciton \#5 is the $2s$ exciton. Furthermore we check that the $2s$ state which is more spread out around the origin in real space than the $1s$ state is more concentrated on the $K$ points in reciprocal space.

Tight-binding calculations lead to quite similar results as can be seen in the second column of Fig. \ref{recip}. The advantage of the TB method is that we can easily obtain the wave functions themselves. It is possible also to have a modulus-phase representation by placing on each point of a grid a circle with color related to the phase between $-\pi$ and $+\pi$ with an opacity proportional to the intensity at this k-point, which is shown in the last columns of Fig.\ \ref{recip}. The four $2p$ states have more rich structures with phases rotating within each triangular spot located at the $K$ points. They can be explained if we use the model recalled above where  the symmetry of the exciton states is governed by the product of the Bloch functions and an envelope function so that the exciton wave function is proportional to $e^{i\bm{K_1}\cdot \bm{r}} g(\bm{r})$ where here $\bm{K_1}$ is one particular $\bm{K}$ vector among the three equivalent ones. In the discrete tight-binding model $\bm{r}$ is the sum of a vector of the triangular lattice and of any first neighbour vector $\bm{\tau}$. Assume now that $g(\bm{r})$ is a $s$ envelope function and only depends on the modulus $r$ of $\bm{r}$. Under a rotation of angle $2\pi/3$, $\bm{K}_1$ is transformed into another equivalent vector modulo a vector $\bm{G}$ of the reciprocal lattice. The exciton state is therefore multiplied by a phase factor equal to $e^{i\,\bm{G}\cdot \bm{\tau}}= \omega$ or $\omega^2$, where $\omega = e^{2i\pi/3}$ is the cubic root of unity, depending on the initial orientation of the lattice with respect to the origin. As mentioned previously, the wave function  transforms as the component $E_+$ of positive chirality of the representation $E$, and the wave function $\Phi_{\bm{k}}$ is given, up to a constant by:
\begin{equation*}
\Phi_{\bm{k}} \propto \int d\bm{r} \; e^{-i\,(\bm{K}_1 -\bm{k}).\bm{r}} g({r}) n(\bm{r}) \; ,
\end{equation*}
where $n(\bm{r})$ is the site density, \textit{i.e.} the sum of Dirac functions on the triangular lattice sites, proportional to $\sum_{\bm{G}} e^{i\bm{G}.(\bm{r}-\bm{\tau})}$.

For $\bm{k}$ close to $\bm{K}_1$, $\bm{k}=\bm{K}_1+\bm{q}$, 
we can neglect the variation of $n(\bm{r})$, \textit{i.e.} keep only the $\bm{G}=0$ term,  and the integral over the angle yields $\Phi_{\bm{k}} \propto \int r dr J_0(qr)g(r)$, where $J_0(x)$ is the Bessel function of zero order. If $g(r)$ is peaked at some average value $\bar{r}$ (remember that $\bar{r}$ is at least equal to the minimum hole-electron distance  $a/\sqrt{3}$), $\Phi_{\bm{k}} \propto J_0(q\bar{r})$ and is therefore peaked at $q=0$ and the intensity is maximum in a circle of radius $\sim 1/\bar{r}$. If $\bm{k}$ is close to another vector, say $\bm{K_2}$, then the integral is multiplied by a factor $e^{-i\bm{G}.\bm{\tau}}$, where $\bm{G}=\bm{K_2}-\bm{K_1}$, \textit{i.e} by a factor  $\omega^2$ or $\omega$. Actually the functions in reciprocal space have he same symmetry properties than in real space.

Assume now that $g(\bm{r})$ has a $p$ symmetry, so that $g(\bm{r}) = g(r) e^{\pm i \varphi_{\bm{r}}}$, where $\varphi_{\bm{r}}$ is the angle between $\bm{r}$ and the $x$-axis. Here also, the integral over this angle can be performed explicitly, so that $\Phi_{\bm{k}} \propto \pm J_1(q\bar{r}) e^{\pm i \varphi_{\bm{q}}}$, where $J_1(x)$ is the Bessel function of order one, and $\varphi_{\bm{q}}$ is now the angle of $\bm{q}$ with the $x$-axis. We conclude that the phase rotates within each circle centred on the $K $ points. This is clearly as shown in Fig.\ \ref{recip}. This proves definitely that the states \#2 to \#4 are of symmetry $p$. More precisely consider first the non degenerate exciton \#3 and \#4. Exciton $\#4$ does show the $A_1$ symmetry already evidenced in real space. Furthermore one can notice that the phases rotates in opposite directions around the $K$ and $K'$ points, and that these rotations are counterbalanced by the rotations between different points of the same family ($K$ or $K'$), in full agreement with the arguments put forward above. The same is true for exciton \#3 except that the amplitudes are odd with respect to the $\sigma_v$ mirrors (phase shift of $\pi$). One can also notice the signature of the $J_1$ Bessel functions: The intensities vanish at the origin of the spots, disappear at larger distances than in the case of $s$ states, and are maximum in between. Actually warping effects along the $\Gamma-M$ lines transform the circles into triangles. The case of the degenerate $E$ state is more complex, since each component seems to mix the behaviours of $s$ and  $p$ states. Mixing between different $E$ states is allowed indeed (this is  also the case for the $2s$ exciton). It is clear however that the main features correspond to rotating phases within circles, and one can check that here the rotations within the circles and those between the circles are in the same direction.

To summarize, although the first excitons considered here are fairly localized, with important lattice effects, they can be classified to some extent within a scheme borrowed from the 2D atomic terminology ($1s, 2s, 2p, \dots$ states) and already applied successfully to TMD. The genuine symmetry of the exciton states is however more precisely related to the representations of the triangular point group. We have also seen that exchange effects are unusually strong for fully invariant states.

\subsection{Fit of the potential}
\label{fit}

\begin{figure}[!ht]
\includegraphics*[width=7.5cm]{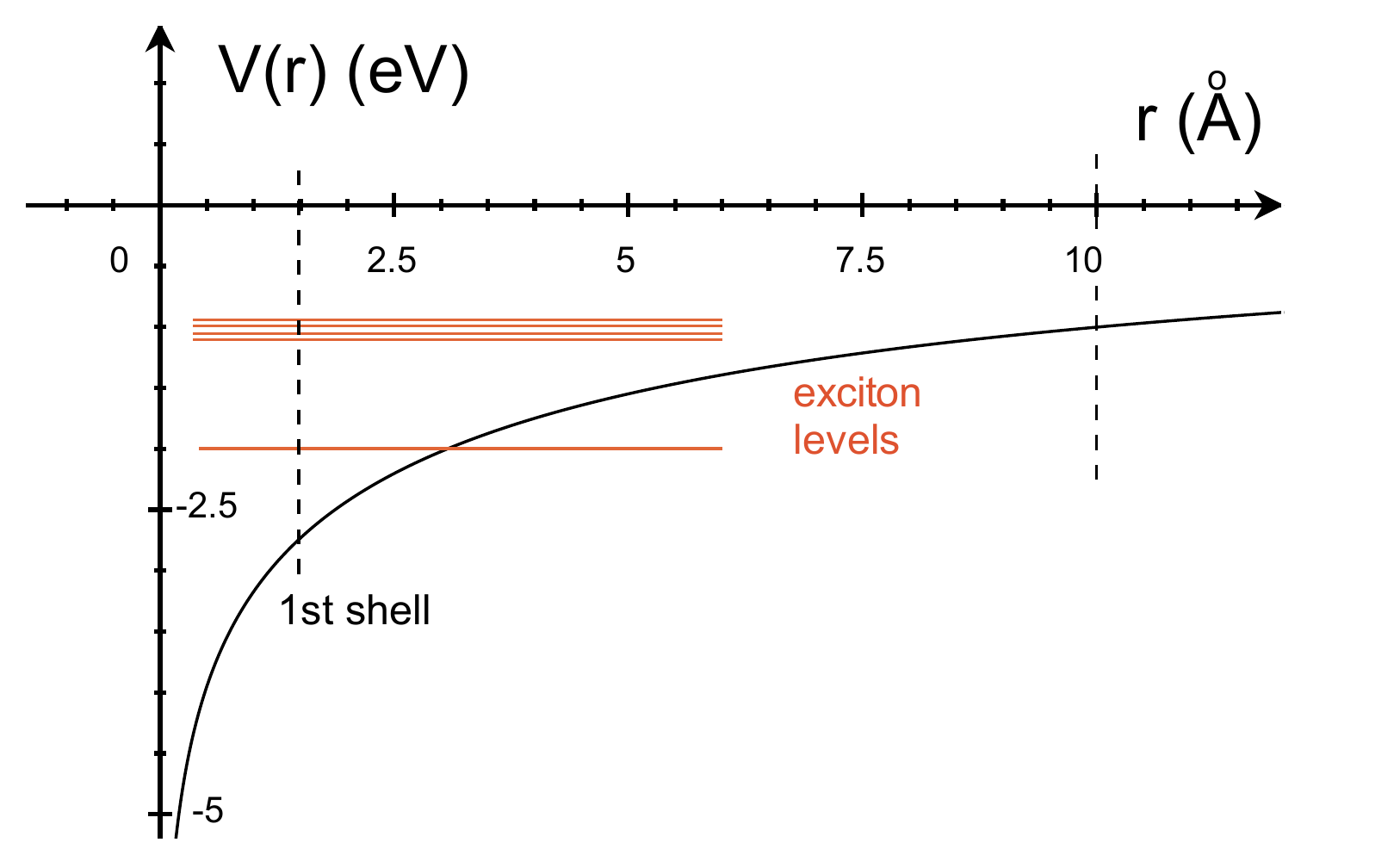} 
\includegraphics*[width=6cm]{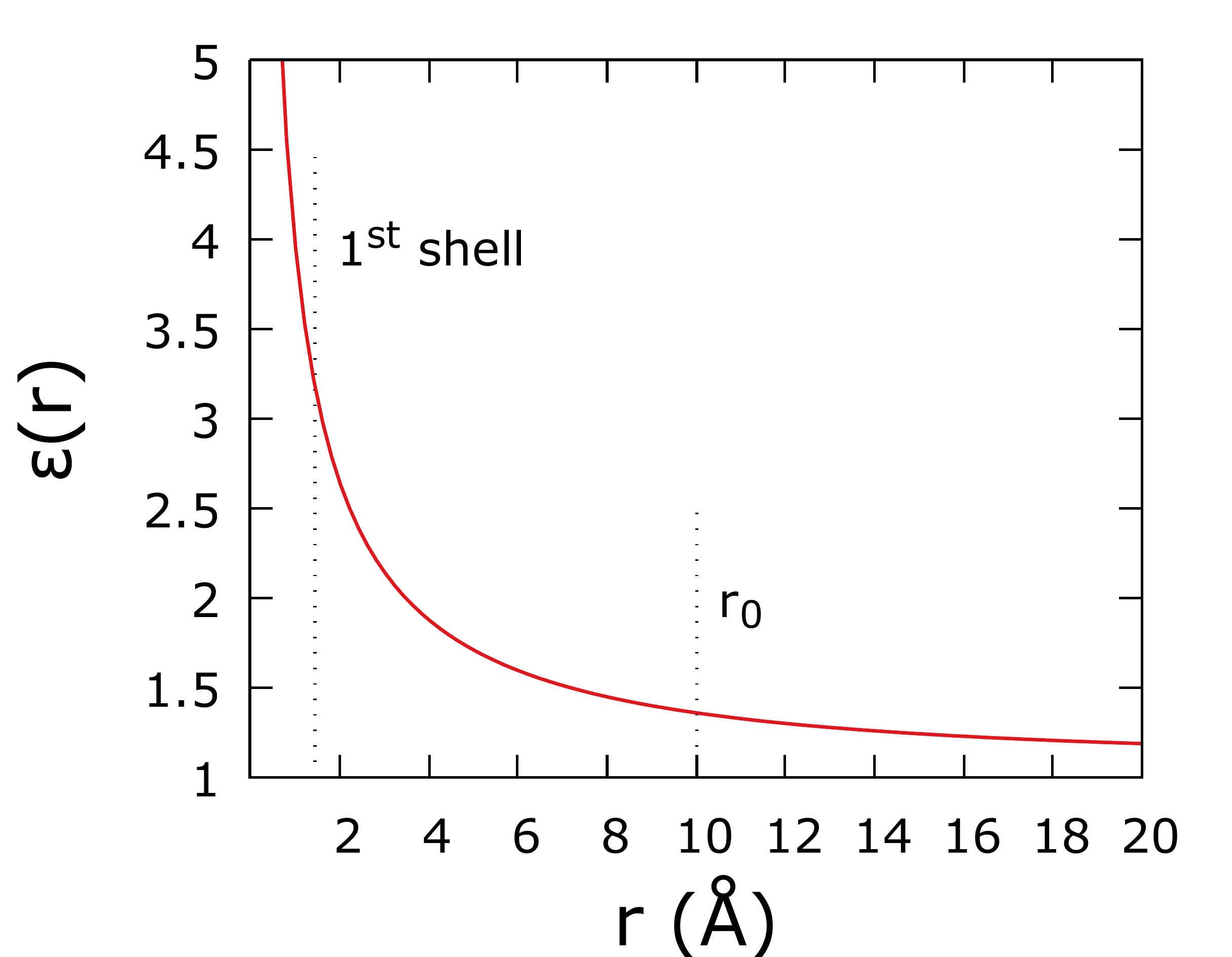}
\caption{Keldysh potential corresponding to the case of hBN-SL; $r_0 = 10.0$ \AA\ and effective dielectric constant $\epsilon(r)$ defined from $V(r)=e^2/\epsilon(r)r$.}
\label{keldysh}
\end{figure}

In a first approach we have tested a  screened Coulomb potential, but it was quickly apparent that it was not possible in this way to reproduce accurately more than the first exciton. Meanwhile several developments in the literature were convincingly arguing that actually it is not possible in 2D to use such a potential and that a genuine 2D electrostatic potential\cite{Keldysh1979} should be used instead.\cite{Cudazzo2011,Berkelbach2013,Chernikov2014,Latini2015,Rodin2014,Wu2015,Pulci2015} We have therefore used the Keldysh potential:
\footnote{More precisely, we have used the simplified form proposed in [\onlinecite{Cudazzo2011}].}
\begin{equation*}
V_{2D}(r)=\frac{\pi e^2}{2r_0} \left[H_0\left(\frac{r}{r_0}\right) - Y_0\left(\frac{r}{r_0}\right)\right] \; ,
\end{equation*}
where the only parameter is the ``screening'' length $r_0$, directly related to the 2D polarisability. Finally, since we are dealing with relative binding energies, our model only depends on two parameters, the excitonic hopping integral $t_{ex} = t^2/\Delta$ and $r_0$. Of course, the Keldysh potential is still defined within a continuous approach which has no reason to apply exactly here where lattice effects are important. In the best fit, the hopping integral $t_{exc} = t^2/\Delta$ is found equal to 1.50 eV, so that $t=2.33$ eV, which is completely consistent with our value $t=2.30$ eV  deduced from \textit{ab initio} band structures. Finally we find $r_0 = 10.0$ \AA .  When $r$ is much larger than $ r_0$ the potential tends to an unscreened Coulomb potential.
Below $r_0$ the potential is screened and becomes logarithmic. Since the first electronic shell around the hole is at a first neighbour B--N distance, about 1.45 \AA , we see in Fig.\ \ref{keldysh} that we are here in the screened regime where the potential is slowly varying.  In Fig.\ \ref{keldysh} we  show the effective distance dependent dielectric constant defined from $V(r) =e^2/\epsilon(r) r$. Notice also that the distance between the two first excitons is completely different from that predicted by the 2D hydrogenic model. This is due in part to these screening effects, but also and more importantly to lattice effects coupled with the specific electronic structure of hBN, since such deviations have already been observed with a fixed dielectric constant (see also Ref. [\onlinecite{Wu2015}]). The main reason is probably that it is nearly forbidden for the hole and the electron to be at the same position, which penalizes principally the binding energy of the ground state exciton.

We have neglected the exchange term here. Actually its short range part contributes generally to the splitting between the spin singlet and triplet states, the latter being dark in principle in the absence of spin-orbit coupling. This is a repulsive (positive effect) absent in the triplet term whose level should therefore be below the singlet one. \textit{Ab initio} calculations predict a splitting about $90$ meV for the $1s$ exciton.\cite{Wirtz2005} In a first order perturbation calculation this splitting is equal to the average of $K^x_{eh}$ in the considered excitonic state. In the case of the ground state exciton, the excitonic wave function is concentrated on the first neighbour shell, so that this splitting is equal to a fraction of $J_{\bm{\tau}}$. As expected then $J_{\bm{\tau}} \simeq 0.1-0.3 $ eV $ \ll U_{\bm{\tau}}$. This is also completely consistent with the variation shown in Table \ref{tabexi} of the exciton energy when suppressing the exchange term. The particular case of the $A_1$ exciton which is very dependent on the intra-atomic values $U_0$ and $J_0$ of the Coulomb and exchange potentials has been discussed above. A simple perturbation method improving the simplest TB model used here can be derived to discuss this effect in more detail and is described in Appendix~\ref{newwannier}.

To summarize, the simple tight-binding model for the excitons in hBN-SL is remarkably successful, even at a quantitative level and the comparison with \textit{ab initio} calculations shows that the effective screened Coulomb potential to be used in a continuous model is really a potential of the Keldysh type in its strongly screened regime.

\section{Optical matrix elements}
\label{optics}

The optical absorption is related to transitions from the ground state (energy $E_\emptyset$) to final states of energy $E_i = E_\emptyset + \hbar\omega$ and therefore to the corresponding matrix elements of the perturbation induced by the electromagnetic field. Each transition $i$ is then characterized by an oscillator strength $f_i = 2m| \langle\emptyset | \bm{v}.\hat{\bm{e}} | i\rangle|^2/\hbar\omega_i$. Here $\hat{\bm{e}}$ is the (unit) vector of  the light polarization, and $\bm{v}$ is the velocity operator.

\subsection{Matrix elements between single particle states}

In the absence of excitonic effets we have just to calculate the matrix elements between valence and conduction Bloch states with identical $\bm{k}$ vectors. It is not difficult to calculate them in the general case,\cite{Margulis2012,Margulis2014,Berkelbach2015} but here we just detail the calculation for states close to the gap where we know that the Bloch functions ``live'' on separate triangular sublattices. Then:
\begin{equation*}
\langle \bm{k}v | \bm{v} |\bm{k}c\rangle = \frac{1}{N} \sum_{\bm{n},\bm{m}} e^{i\bm{k}(\bm{m}-\bm{n})} \langle \bm{n}A | \bm{v} |\bm{m}B\rangle\; .
\end{equation*}
There is no unique way of calculating the matrix elements of $\bm{v}$, depending on whether we express it using the momentum operator\cite{Jorio2011} or the relation $\bm{v}= [\bm{r},H]/i\hbar$. Both methods are equivalent in an exact treatment but not when using an incomplete basis as is the case of our tight-binding basis. The  second method has the disadvantage to use the $\bm{r}$ operator which is not always well defined in periodic systems, but here there is no problem and the advantage is to work directly in real space, assuming that $\bm{r} |\bm{n}\rangle \simeq \bm{n} |\bm{n}\rangle $, and therefore:
\begin{equation}
\langle \bm{n}A | \bm{v} |\bm{m}B\rangle =-\frac{1}{i\hbar}(\bm{m}-\bm{n}) t \; ,
\label{velocity}
\end{equation}
if $\bm{m}$ and $\bm{n}$ are first neighbours (on the honeycomb lattice), and zero otherwise. Then:
\begin{align*}
\langle \bm{k}v | \bm{v} |\bm{k}c\rangle &= \frac{1}{N} \sum_{\bm{n},\bm{m}} e^{i\bm{k}(\bm{m}-\bm{n})} \frac{it}{\hbar} (\bm{m}-\bm{n})
= \frac{it}{\hbar}\sum_{\alpha}e^{i\bm{k}.\bm{\tau_\alpha}}  \bm{\tau_\alpha}\\
&=\frac{it}{\hbar}\nabla_{\bm{k}} \gamma(\bm{k}) \; .
\end{align*}
In the limit $\bm{k}\to K$, one finds $\nabla_{\bm{k}} \gamma(\bm{k}) \simeq -\frac{3}{2}ia(\hat{\bm{x}} + i \hat{\bm{y}})$, where $a$ is here the nearest neighbour distance, \textit{i.e.} the lattice parameter divided by $\sqrt{3}$ and $\hat{\bm{x}}$ and $\hat{\bm{y}}$ are the unit vectors along the $x$-axis and the $y$-axis, respectively, so that finally:
\begin{equation*}
\langle \bm{K}v | \bm{v}.\hat{\bm{e}} |\bm{K}c\rangle|  \simeq v |e_x + i e_y| \quad;\quad \hbar v=3a t/2 \; .
\end{equation*}
Notice that $\Delta/v^2$ is nothing but the effective mass $m^*$ of the conduction and valence bands at point $K$. Using $\Delta \simeq 3$ eV and $v \simeq 1$ km/s (as the Fermi velocity of graphene precisely given by $3at/2\hbar$), we obtain $m^*/m \simeq 0.54$.

At point $K'$, $e_x + i e_y$ is replaced by $e_x - i e_y$. The matrix elements are maximum for circular polarized light and, for linearly polarized light, the matrix element is constant and equal to $v$.  The oscillator strength is equal to $(m/m^*)(2\Delta/\hbar\omega)$ when $\hbar\omega$ is larger than the gap, \textit{i.e.} about 2 close to the gap (equal to $2 \Delta$). The absorption, proportional to $\sum_i f_i/\hbar\omega_i$, is therefore proportional to the density of states of the triangular lattice divided by $(\hbar\omega)^2$. Its shape is characterized by a discontinuity at the edge and a Van Hove singularity at a distance equal to $t_{exc}$ above the edge, as shown in Fig.\ \ref{triangle}.

\subsection{Matrix elements between excitonic states}

In the presence of excitons we have now to calculate the matrix element $\langle \emptyset | \bm{v}.\hat{\bm{e}} |\Phi\rangle$, where $|\Phi\rangle$ is the exciton state.
From \eqref{R} and \eqref{velocity}, we see that:

\begin{equation*}
\bm{v}|\emptyset\rangle = \frac{-t}{i\hbar} \sum_{\bm{n},\bm{m}}{}' (\bm{m}-\bm{n})a_{\bm{m}B}^+ a_{\bm{n}A}^{} \ket{\emptyset}
= \frac{it}{\hbar} \sqrt{N} \sum_{\alpha} \bm{\tau}_\alpha|\bm{\tau}_\alpha\rangle  \; ,
\end{equation*}
so that:
\begin{equation*}
\langle \emptyset | \bm{v}.\hat{\bm{e}} |\Phi\rangle = \frac{-it}{\hbar} \sqrt{N} \sum_{\alpha} \hat{\bm{e}}.\bm{\tau}_\alpha \langle \bm{\tau}_\alpha|\Phi\rangle
\end{equation*}
Defining the dipole $\bm{d}_{\Phi}$ associated with the exciton $\Phi$ through:
\begin{equation*}
\bm{d}_{\Phi} =  \sum_{\alpha} \bm{\tau}_\alpha \langle \bm{\tau}_\alpha|\Phi\rangle  \;,
\end{equation*}
we see that $|\langle \emptyset | \bm{v}.\hat{\bm{e}} |\Phi\rangle| = (t\sqrt{N}/\hbar) |\hat{\bm{e}}.\bm{d}_\Phi |$. Thus only the local components of the exciton wave fonction contribute to the optical matrix element. This is completely equivalent to the  statement that, within the usual hydrogenic model, only  $s$ states  contribute (Elliott theory, see [\onlinecite{Yu2010,Toyozawa2003}]). Here we have an equivalent selection rule: the dipole $\bm{d}_\Phi$ should not vanish; in particular the wave function $\langle\bm{R}|\Phi\rangle$ should have finite components on the first neighbours $\bm{R}=\bm{\tau}_\alpha$.

\subsection{Application to the five first excitons of hBN-SL}

Consider first the ground state $E$ exciton. It has two components $\Phi^+$ and $\Phi^-$ which can be chosen as those corresponding to circular polarizations, so that the components are the cubic roots of unity, $ \langle\bm{\tau}_\alpha|\Phi^\pm\rangle = C_\Phi e^{\pm\frac{2i\pi}{3}(\alpha-1)}$, and  $ \hat{\bm{e}}.\bm{d}_\Phi = -\frac{3}{2} a\, C_\Phi (e_x\pm ie_y)$, and finally $|\langle \emptyset | \bm{v}.\hat{\bm{e}} |\Phi\rangle|/\sqrt{N} = C_\Phi v |e_x+ ie_y|$. Here, $C_\Phi$ is the amplitude of the exciton state on the first neighbours, at most equal to $1/\sqrt{3}$. In the case of single particle transitions the oscillator strength was of the order of $mv^2/\Delta$ for a transition close to the gap; hence a total oscillator strength of the order of $N$ times this value. We see here that the oscillator strength of the exciton is of the same order of magnitude if $C_\Phi$ is large, \textit{i.e.} if the exciton is strongly localized, which is the case here. Actually from \textit{ab initio} calculations, the weight $C_\Phi^2$ is found about  one third its maximum value 1/3. In other words 30\% of the weight of the ground state is concentrated on the first triangular shell. TB calculations on the other hand find a weight about 50\%. 
The oscillator strength of the other excitons are smaller. The second exciton as well as the last one  (\#5) has the same symmetry as the first one but their amplitude on the first neighbours is weak. Finally the two other ones studied above  (\#3-4) are dark because their symmetry are characterized by representations $A_1$ and $A_2$ different from the vectorial representation $E$, so that $\bm{d}_\Phi = 0$ and this is confirmed by the \textit{ab initio} calculations. Finally the ground state exciton  takes almost all the oscillator strength.

\section{Discussion}
\label{discussion}

The excitons of hBN-SL have been characterized in detail. The first one, of lowest energy is particularly localized. Is it a Frenkel or Wannier-Mott exciton ? This discussion is somewhat semantic. It is in some sense similar to the long standing debate between the Heitler-London (atomic) approach and the Hund-Mulliken (``molecular'') approach to single particle properties. In practice, it turns out that in solids the band Hund-Mulliken approach is more fruitful since it can deal with many situations except when correlations effects are very strong. Even then, specific approaches ``\`a la Hubbard'' can be used and compete with the methods of quantum chemistry (interaction configuration approach). In between, the tight-binding method has proven very efficient to deal with electrons sharing itinerant properties (conductivity) and localized ones (magnetism, chemical bonding). We are certainly here in a similar situation. The localized excitons of hBN-SL can be described within a TB-Wannier framework, but  cannot be described accurately within a $\bm{k}\cdot \bm{p}$ approach similar to the nearly free electron approach of electronic properties. They could simply be described as ``tightly-bound excitons''. We have shown indeed that in the case of hBN-SL which is a genuine case study, the tight-binding approach can be very accurate by fitting to \textit{ab initio} data a few parameters. 

On the experimental side optical properties of hBN-SL are not available yet, but there are already some indications that the expected main exciton is observed. In the case of bulk hBN, stacking effects induce splittings of this exciton level which are observed. Progress in the analysis of these stacking effects are under progress. Finally dispersion effects as well as exciton-phonon coupling remain to be studied.

\textit{Note added}. A recent paper presents a model for excitons in dichalcogenides whose spirit is quite similar to our tight-binding model.\cite{Gunlycke2016}

\begin{acknowledgments}
 A. Molina-S\'anchez and L. Wirtz acknowledge support from the National Research Fund, Luxembourg (Projects C14/MS/773152/FAST-2DMAT and INTER/ANR/13/20/NANOTMD). F. Ducastelle and H. Amara are indebted to L. Schu\'{e}, J. Barjon, and A. Loiseau for numerous fruitful discussions. The research leading to these results has received funding from the European Union Seventh Framework Programme under grant agreement no. 604391 Graphene Flagship. We acknowledge funding by the French National Research Agency through Project No. ANR-14-CE08-0018.
\end{acknowledgments}

\appendix

\section{A very simplified model for the ground state exciton}
\label{greenf}

The ground state  exciton is so localized that its properties do not depend too much on the long range part of the potential. It is useful therefore to examine the properties of a model where the range of the potential is limited to the three first neighbours of the central hole. We have then to diagonalize the following hamiltonian:

\begin{align*}
H_{eh} &= H^0_{eh} + U \\
\bra{\bm{R}}H^0_{eh}\ket{\bm{R'}} &=  3t_{exc}\quad \mbox{if}\quad  \bm{R} =\bm{R}' \\
&= t_{exc}\quad \mbox{if} \quad \bm{R} \quad \mbox{et} \quad \bm{R}' \quad \mbox{are first neighbours}\\
&=0\quad  \mbox{otherwise}, \\
U &= \sum_{\bm{R}=1,2,3} \ket{\bm{R}} u \bra{\bm{R}} \; ,
\end{align*}
where the three sites surrounding the hole at the origin are labelled $1,2,3$. 

\subsection{Green functions}

The resolvant or Green function corresponding to this hamiltonian is $G(z)=(z-H_{eh})^{-1}$. $G^0(z)=(z-H^0_{eh})^{-1}$ is then the Green function of the triangular lattice. With the chosen origin of energies, the spectrum of $H^0_{eh}$ starts at $E=0$ with a jump equal to $\pi\sqrt{3} \,t_{exc}$. In the presence of the attractive potential $u$ we can have a bound state if the determinant of the operator $(1-G^0U)$ within the space of dimension 3 generated by the three states $|1\rangle,|2\rangle$ and $|3\rangle$ vanishes. Let now $F_0$ and $F_1$ be the diagonal and off-diagonal matrix elements of $G^0$, respectively $F_0  = \langle 1|G^0|1\rangle = \langle 2|G^0|2\rangle = \langle 3|G^0|3\rangle$; $F_1 = \langle 1|G^0|2\rangle = \langle 2|G^0|3\rangle = \langle 3|G^0|1\rangle$. We find a double solution $(F_0 -F_1)=1/u$ and a simple solution $(F_0 +2F_1)=1/u$. Using standard methods one can determine the behaviour of $F_0$ and $F_1$ close to the origin. First, assuming a constant density of states $n(E) \simeq 1/W$, we find that: 
\begin{equation*}
F_0 = \frac{1}{W} \log\frac{z}{z-W}  \; ,
\end{equation*}
so that $F_0(E<0)\simeq \frac{1}{W} \log (|E|/W)$ when $E$ is close to 0. Similarly, $F_1$ is found to behave as $-F_0/2$. Then $F_0-F_1$ is negative and diverges logarithmically below $E=0: F_0-F_1 \simeq (3/2W) \log (|E|/W)$ whereas $F_0+2F_1$ tends to a constant. As a result the first equation has always a negative solution $E_{exc}$ for $E$, solution such that $F_0(E)-F_1(E) = -1/|u|$. With the previous model for $F_0$ and $F_1$, we obtain, for small values of $|u|$, $|E_{exc}|/W = \simeq\exp(-{\frac{2W}{3|u|}}).$ If $|u|$ is large, $ E_{exc} \simeq u$.
The corresponding eigenstates are, as expected, the ``chiral''states $| \phi^+\rangle \propto |1\rangle + \omega |2\rangle +\omega^2 |3\rangle$ and  $ |\phi^\pm\rangle 
\propto |1\rangle + \omega^2 |2\rangle +\omega |3\rangle$, where $\omega$ is the cubic root of unity, $\omega=e^{2i\pi/3}$. We recover our exciton of symmetry $E$. The components of $| \Phi ^\pm\rangle$ beyond the (1,2,3) cluster can be obtained from the equation $| \phi^\pm\rangle = G^0U | \phi^\pm\rangle$, \textit{i.e.} $\langle\bm{R} | \phi^\pm\rangle = \sum_{\bm{R}'=1,2,3}\langle\bm{R} |G^0 |\bm{R}' \rangle u  \langle\bm{R}' | \phi^\pm\rangle$.\\

\subsection{Reciprocal space}
\begin{figure}[!ht]
\includegraphics*[width=5cm]{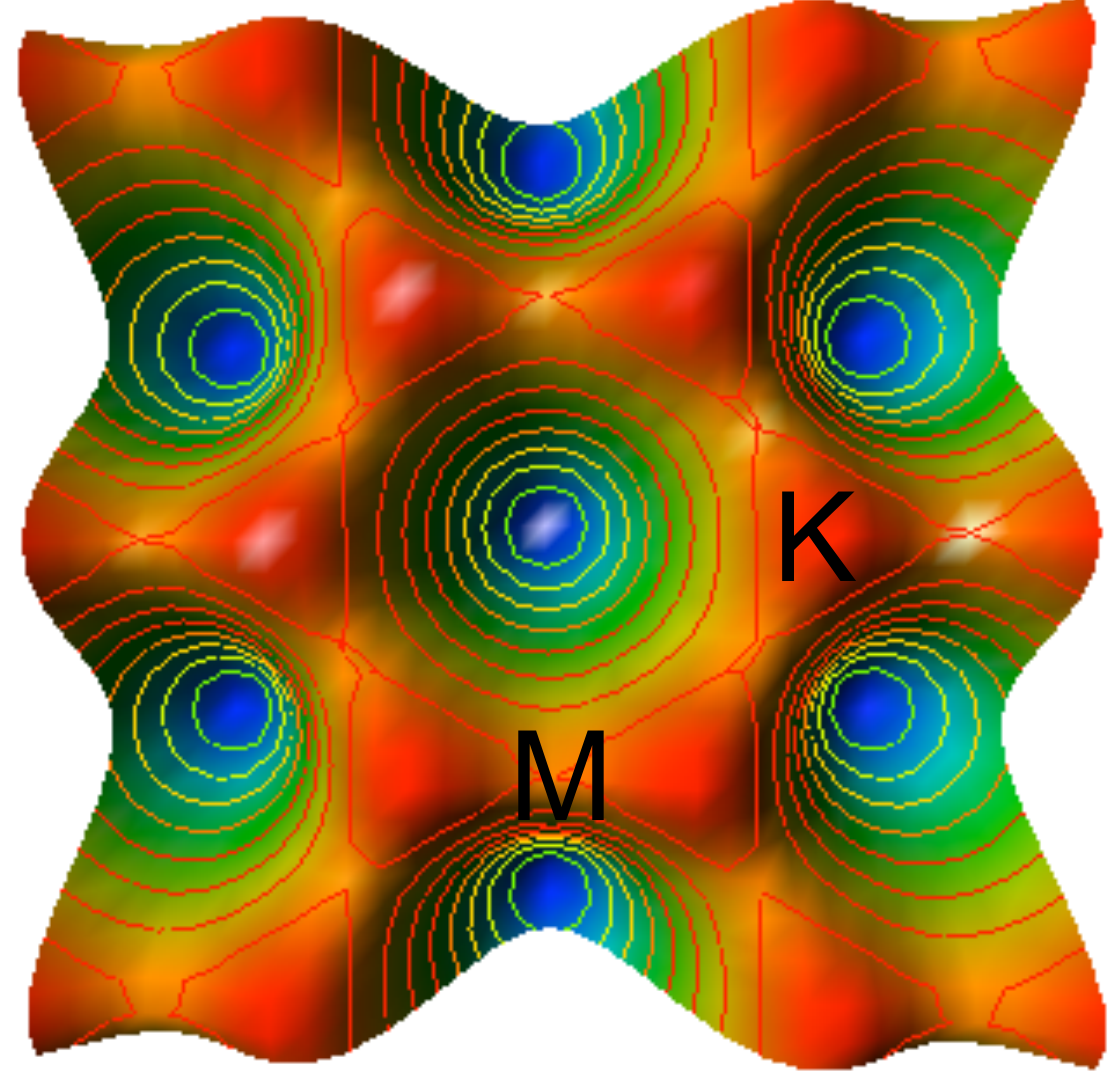} 
\caption{Tight-binding weight of the ground state exciton in reciprocal space.}
\label{exc_recip}
\end{figure}

We can also express $|\phi^\pm\rangle$ in reciprocal space:
\begin{align*}
\Phi^\pm_{\bm{k}} &=  \langle\bm{k}|\phi^\pm\rangle =\frac{1}{\sqrt{N}}\sum_{\bm{R}}e^{-i\bm{k}.\bm{R}} \langle\bm{R} | \phi^\pm\rangle \\
& \simeq \frac{1}{\sqrt{3N}} \gamma(\pm\bm{K}-\bm{k}))    \; ,	\\
\end{align*}
where we have limited the sum to the first neighbours and taken into account that $e^{i\bm{K}.\bm{R}} = 1, \omega, \omega^2$ when $\bm{R} = 1, 2, 3$ provided $K$ is chosen along the $x$-axis, as in Fig.\ \ref{notations}. Thus, up to a normalization constant the weight $|\Phi^\pm_{\bm{k}}|^2$ is equal to $|\gamma({\bm{k}\mp\bm{K}})|^2$. Since $|\gamma(\bm{k})|^2$ is maximum when $\bm{k} = 0$, we see that $|\Phi^+_{\bm{k}}|^2$ and $|\Phi^-_{\bm{k}}|^2$ are peaked at points $K$ and $K'=-K$, respectively. As expected the sum is maximum on the boundary of the Brillouin zone, as shown in Fig.\ \ref{exc_recip}.

\section{Exchange contribution}
\label{echange}
The exchange contribution involves integrals of type:
\begin{widetext}
\begin{equation*}
+ \int d\bm{r} d\bm{r}' \varphi_v(\bm{r}-\bm{R}_n)   \varphi_v(\bm{r}'-\bm{R}_p)  \frac{2}{|\bm{r}-\bm{r}'|} 
 \varphi_c(\bm{r}-\bm{R}_m)  \varphi_c(\bm{r}'-\bm{R}_q)   \; .
\end{equation*}
\end{widetext}
The largest integrals correspond to cases where the overlap is minimum for $\bm{r}$ and $\bm{r}'$ integrations. But since we have forbidden site coincidence for valence and conduction orbitals, the best we can do is to consider first neighbour overlap between $\varphi_v(\bm{r}-\bm{R}_n) $ and  $\varphi_c(\bm{r}-\bm{R}_m)$ and between $\varphi_v(\bm{r}'-\bm{R}_p) $ and  $\varphi_c(\bm{r}'-\bm{R}_q)$. Therefore we only keep the terms $\bm{R}_m = \bm{R}_n + \bm{\tau}$ and $\bm{R}_q = \bm{R}_p + \bm{\tau}'$, where $\bm{\tau}$ and $\bm{\tau}'$ are first neighbours on the honeycomb lattice.
The largest terms occur when $n=p$, and the integral becomes a function $J(\bm{\tau},\bm{\tau}\,'\bm{\rho)}$ of $\bm{\tau}, \bm{\tau}'$,  and $\bm{\rho}$ where $\bm{\rho}$ measures the separation between the pairs $\bm{\tau}$ and $\bm{\tau}' :$

\begin{widetext}
\begin{equation*}
J(\bm{\tau},\bm{\tau}\,'\bm{\rho)} 
= \int d\bm{r} d\bm{r}\,'
\varphi_v(\bm{r}) \varphi_c(\bm{r}-\bm{\tau}) 
\frac{2}{|\bm{r}-\bm{r}\,'|}
\varphi_v(\bm{r}\,'-\bm{\rho})
\varphi_c(\bm{r}\,'-\bm{\rho}-\bm{\tau}\,')  \; ,
\end{equation*}
\end{widetext}
which induces in the tight-binding hamiltonian an effective overlap integral $\langle \bm{\tau} | K^x | \bm{\tau}'\rangle$  which depends on the first neighbours of the origin $\bm{\tau}$ and $\bm{\tau}'$:
\begin{equation*}
\langle \bm{\tau} | K^x_{eh} | \bm{\tau}'\rangle =
 \sum_{\bm{\rho}} J(\bm{\tau},\bm{\tau}\,'\bm{\rho)} 
\end{equation*}
To lowest order, when  $\bm{\tau} = \bm{\tau}'$,  this adds a local term $J_{\bm{\tau}}$ to the direct term on the first neighbours:
\begin{widetext}
\begin{equation*}
J_{\bm{\tau}} = J(\bm{\tau},\bm{\tau},\bm{\rho=0} ) = \int d\bm{r} d\bm{r}\,'
\varphi_v(\bm{r}) \varphi_c(\bm{r}-\bm{\tau}) 
\frac{2}{|\bm{r}-\bm{r}\,'|}
\varphi_v(\bm{r}\,')
\varphi_c(\bm{r}\,'-\bm{\tau})  \; .
\end{equation*}
\end{widetext}
On the other hand when we sum all contributions corresponding to all separations of the distant  $\bm{\tau}$ and $\bm{\tau}'$ pairs we obtain a sum of dipolar contributions which are known to be singular here ($\bm{Q} \to 0$ limit). In 3D this produces the so-called longitudinal-transversal splitting.\cite{Denisov1973,Toyozawa2003} In 2D the singularity is weaker, with terms varying as $|\bm{Q}|$.\cite{Wu2015,Qiu2015} This will be discussed elsewhere.

\begin{figure}[!ht]
\includegraphics*[width=5cm]{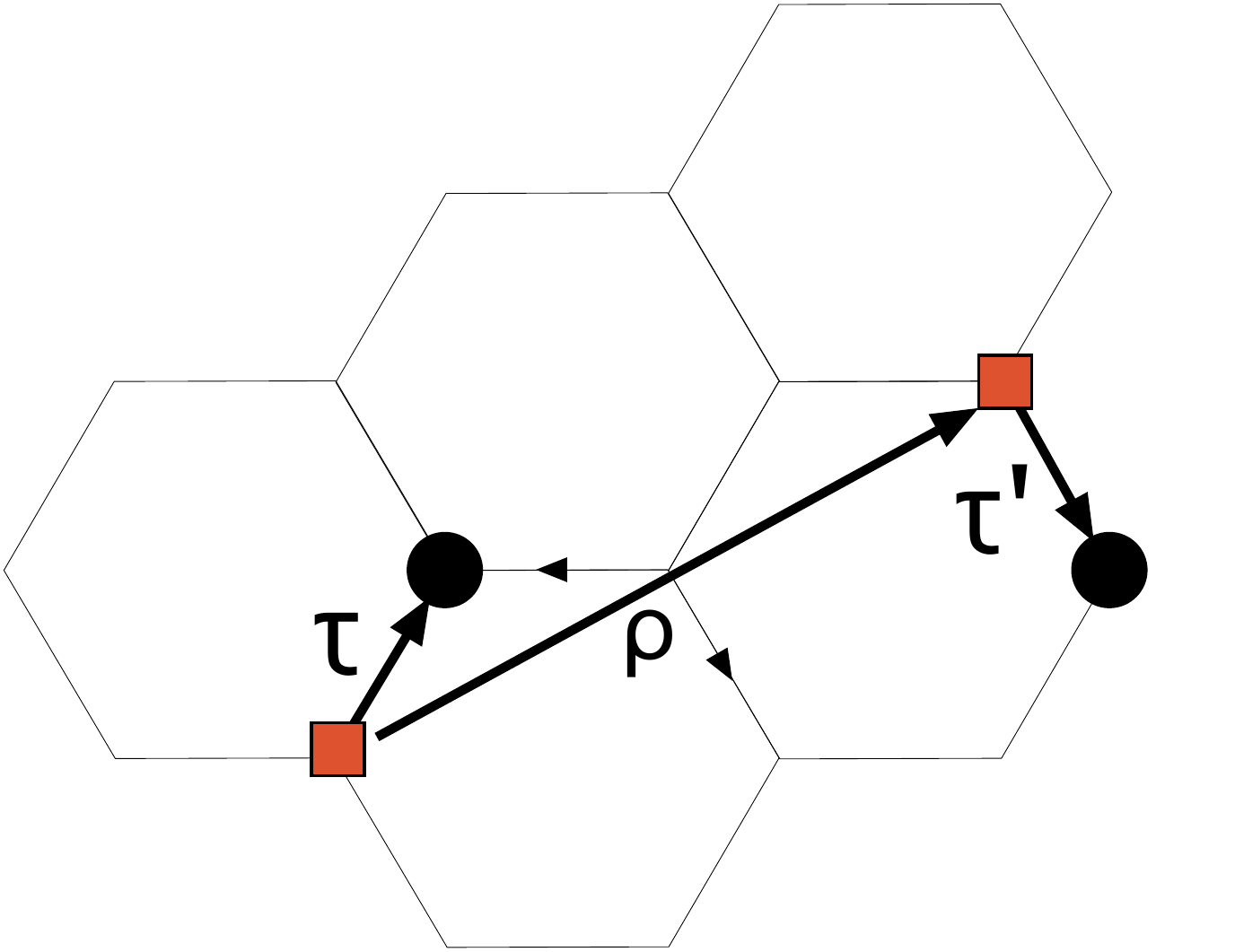} 
\caption{Schematic representation of the geometry of the integral $J(\bm{\tau},\bm{\tau}\,'\bm{\rho)}$. Hole positions are shown as red squares and electron ones as black circles.}
\label{jtau}
\end{figure}

\section{An improved Wannier model}
\label{newwannier}
The tight-binding model developed in the main text is based on several approximations. To lowest order in $t/\Delta$ the Wannier functions corresponding to the valence and conduction bands are taken as the $\pi$ orbitals on the nitrogen and boron sites, respectively. Then the kinetic energy part of the exciton hamiltonian is approximated by its second order term in $t/\Delta$ and finally the Coulomb matrix elements $U_{\bm{R}}$ are calculated using atomic orbitals, \textit{i.e.} Wannier function of zero order. We will see that higher order terms induce corrections of order $(t/2\Delta)^2 \simeq 0.1$. This is not negligible but has been implicitly taken into account via our fitting procedure for the interactions where $\bm{R}\neq 0$ on the triangular lattice. Problems arise because, to higher order, other interactions become allowed, in particular when $\bm{R} = 0$. Let us then define more accurate Wannier state $|\bm{n}\pm\rangle_w$ from the exact eigenstates $|\bm{k}\pm\rangle$ defined in Eq. \eqref{fpropres}. We choose the phases such that these Wannier states  reduce to the atomic states when $t \to 0$. Then, to linear order in $t/\Delta$:
\begin{equation}
\begin{split}
|\bm{m}+\rangle_w &= \frac{1}{\sqrt{N}} \sum_{\bm{k}} e^{-i.\bm{k}.\bm{m}} |\bm{k}+\rangle \\
&\simeq |\bm{m}B\rangle -\frac{t}{2\Delta}\sum_{\bm{\tau}}	|\bm{m}B - \bm{\tau}\rangle	\\
|\bm{n}-\rangle_w &= \frac{1}{\sqrt{N}} \sum_{\bm{k}} e^{-i.\bm{k}.\bm{n}} |\bm{k}-\rangle  \\
& \simeq |\bm{n}A\rangle +\frac{t}{2\Delta}\sum_{\bm{\tau}}	|\bm{n}A +\bm{\tau}\rangle	\; .\\
\end{split}
\end{equation}
The sites $\bm{n}A(\bm{B})$ are on the $A(B)$ sublattices and the sites $\bm{m}B - \bm{\tau}$ $(\bm{n}A + \bm{\tau})$ are on the $A(B)$ sublattices. These Wannier states remain centred on boron $(B)$ and nitrogen $(A)$ sites, respectively, so that we can continue to use sublattice labels $A,B$ instead of band labels $\pm$~, but they spread on the neighbouring sites, on the other sublattices. The  excitonic kinetic energy term calculated to second order in $t/\Delta$ has exactly the form derived previously, but we are now able to calculate the corrections to the Coulomb term. The Wannier fonctions on neighbour sites overlap so that the Coulomb matrix elements between electron and hole Wannier states labelled by $\bm{n}A$ and $\bm{m}B$ involve not only the usual Coulomb integrals $U_{\bm{nm}}$ but also integrals involving sites  $\bm{n}A+\bm{\tau}$ and $\bm{m}B-\bm{\tau}$. More precisely, let us define the Wannier electron-hole states $|\bm{R}\rangle_w$:
\begin{equation*}
|\bm{R}\rangle_w =  \frac{1}{\sqrt{N}} \sum_{\bm{n}} w^+_{\bm{n}A+\bm{R}} \; w_{\bm{n} A}^{}  \ket{\emptyset} \; ,
\end{equation*}
where $w^+_{\bm{m}}$ is the creation operator in the Wannier state $|\bm{m}B\rangle_w \equiv |\bm{m}+\rangle_w $, and $w_{\bm{n}}$ is the destruction operator in the Wannier state $|\bm{n}A\rangle_w \equiv |\bm{n}-\rangle_w $. $\bm{R}$ is as previously a vector between the two sublattices.
The matrix element of the direct Coulomb kernel now becomes:
\begin{equation*}
_w\langle\bm{R}|K_{eh}^d|\bm{R'}\rangle_w \simeq U_{\bm{R}} \,\delta_{\bm{R},\bm{R}'}  + \frac{t^2}{2\Delta^2} \, \sum_{\bm{\tau},\bm{\tau}'}\delta_{\bm{R}-\bm{\tau} ,\bm{R}'-\bm{\tau}' } U_{\bm{R}-\bm{\tau}} \, .
\end{equation*}
We will keep only the corrective terms involving $U_0$. Thus these  terms only correct the  matrix elements between neighbours of the origin:
\begin{equation*}
_w\langle\bm{\tau}|K_{eh}^d|\bm{\tau}'\rangle_w \simeq U_{\bm{\tau}} \,\delta_{\bm{\tau},\bm{\tau}'} + \frac{t^2}{2\Delta^2} \,U_0 \; .
\end{equation*}
Actually, second order terms in the  Wannier function expansions also contribute, but they do not involve $U_0$. This perturbation expansion is the counterpart in real space of the developments in reciprocal space (and within the $\bm{k}.\bm{p}$ approximation), performed in [\onlinecite{Wu2015}], [\onlinecite{Srivastava2015}], and [\onlinecite{Zhou2015}]. 

Although $U_0$ cannot be derived from the continuous Keldysh potential, it should be significantly larger than its value at the first neighbour positions, about 3 eV (see Fig.\ \ref{keldysh}), and the perturbation is not negligible \textit{a priori}. Let us estimate to lowest order the correction $\delta E$ to the energy of the excitonic state $|\Phi\rangle$:
\begin{equation*}
\delta E \simeq \langle\Phi |\delta K_{eh}^d|\Phi\rangle =  \frac{t^2}{2\Delta^2} \,U_0 |\sum_{\bm{\tau}}\langle\bm{\tau}|\Phi\rangle|^2 < 0 \; .
\end{equation*}
The sum of the amplitude on the first shell, $\sum_{\bm{\tau}}\langle\bm{\tau}|\Phi\rangle$ is non vanishing only for states of full symmetry $A_1$. This is the case of the exciton \#4 discussed in the main text. As discussed there however, the exchange contribution has also  to be taken into account and the ``corrective'' term  proportional to $2J_0$ (the calculation is similar to that derived above for the Coulomb term) is here more important than the usual $J_{\bm{\tau}}$ (see Appendix \ref{echange}). Since $2J_0$ is a positive contribution at least equal to $|U_0|$, there is a compensation effect and the global correction to our TB model should be slightly positive, which is consistent with the results shown in Table \ref{tabexi}. On the other hand this discussion shows that neglecting exchange effects, which is common practice, is not valid here for hBN when dealing with fully symmetric excitons. In this case the singulet-triplet splitting is huge as can be seen in the same Table.


%

\end{document}